%% file: m4opt.tex
\documentclass[twocolumn,twocolappendix,times]{aastex631}
\usepackage{acronym}
\usepackage{amsmath}
\usepackage{mathtools}

\received{2025 February 2}
\revised{2025 April 8}
\accepted{2025 April 23}
\published{2025 July 14}
\submitjournal{PASP}
\shorttitle{Optimal Follow-Up of \acs{GW} Events with \acs{UVEX}}
\shortauthors{Singer et al.}

\acrodef{ACCESS}[ACCESS]{Advanced Cyberinfrastructure Coordination Ecosystem: Services and Support}
\acrodef{ALMA}[ALMA]{Atacama Large Millimeter/Submillimeter Array}
\acrodef{BBH}[BBH]{binary black hole}
\acrodef{BNS}[BNS]{binary neutron star}
\acrodef{CDF}[CDF]{cumulative distribution function}
\acrodef{CSR}[CSR]{concept study report}
\acrodef{ETC}[ETC]{exposure time calculator}
\acrodef{FOR}[FOR]{field of regard}
\acrodefplural{FOR}[FORs]{fields of regard}
\acrodef{FOV}[FOV]{field of view}
\acrodefplural{FOV}[FOVs]{fields of view}
\acrodef{GSFC}[GSFC]{Goddard Space Flight Center}
\acrodef{GRB}[GRB]{gamma-ray burst}
\acrodef{GW}[GW]{gravitational wave}
\acrodef{HEALPix}[HEALPix]{Hierarchical Equal Area isoLatitude Pixelization}
\acrodef{HQ}[HQ]{headquarters}
\acrodef{ICRS}[ICRS]{International Celestial Reference System}
\acrodef{ITRS}[ITRS]{International Terrestrial Reference System}
\acrodef{KN}[KN]{kilonova}
\acrodefplural{KN}[KNe]{kilonovae}
\acrodef{LCO}[LCO]{Las Cumbres Observatory}
\acrodef{LIGO}[LIGO]{the Laser Interferometer Gravitational-Wave Observatory}
\acrodef{LP}[LP]{linear programming}
\acrodef{MWC}[MWC]{maximum weighted coverage problem}
\acrodef{MIDEX}[MIDEX]{Medium-Class Explorer}
\acrodef{MILP}[MILP]{mixed integer linear programming}
\acrodef{MoO}[MoO]{Mission of Opportunity}
\acrodef{M4OPT}[M$^4$OPT]{the Multi-Mission Multi-Messenger Observation Planning Toolkit}
\acrodef{NCSA}[NCSA]{the National Center for Supercomputing Applications}
\acrodef{NPR}[NPR]{NASA Procedural Requirements}
\acrodef{NOSA}[NOSA]{NASA Open Source Agreement}
\acrodef{NS}[NS]{neutron star}
\acrodef{NSBH}[NS--BH]{neutron star--black hole}
\acrodef{NSF}[NSF]{the U.S. National Science Foundation}
\acrodef{NUV}[NUV]{near ultraviolet}
\acrodef{FUV}[FUV]{far ultraviolet}
\acrodef{PSF}[PSF]{point spread function}
\acrodef{ROI}[ROI]{region of interest}
\acrodef{SADA}[SADA]{solar array drive assembly}
\acrodef{SN}[SN]{supernova}
\acrodefplural{SN}[SNe]{supernovae}
\acrodef{S/N}[S/N]{signal to noise ratio}
\acrodef{SDSC}[SDSC]{the San Diego Supercomputing Center}
\acrodef{STIS}[STIS]{Space Telescope Imaging Spectrograph}
\acrodef{SV}[SV]{site visit}
\acrodef{ToO}[ToO]{target of opportunity}
\acrodefplural{ToO}[ToOs]{targets of opportunity}
\acrodef{ULTRASAT}[ULTRASAT]{the Ultraviolet Transient Astronomy Satellite}
\acrodef{UV}[UV]{ultraviolet}
\acrodef{UVEX}[UVEX]{the UltraViolet EXplorer}
\acrodef{ZTF}[ZTF]{Zwicky Transient Facility}

\begin{document}

\title{Optimal Follow-Up of Gravitational-Wave Events with \acf{UVEX}}

\author[0000-0001-9898-5597]{Leo P. Singer}
\affiliation{Astroparticle Physics Laboratory, NASA Goddard Space Flight Center, Greenbelt, MD 20771, USA}
\affiliation{Joint Space-Science Institute, University of Maryland, College Park, MD 20742, USA}
\email{leo.p.singer@nasa.gov}

\author[0000-0002-9225-7756]{Alexander W. Criswell}
\affiliation{Minnesota Institute for Astrophysics, University of Minnesota, Minneapolis, MN 55455, USA}
\affiliation{School of Physics and Astronomy, University of Minnesota, Minneapolis, MN 55455, USA}
\affiliation{Department of Physics and Astronomy, Vanderbilt University, Nashville, TN 37240, USA}
\affiliation{Department of Life and Physical Sciences, Fisk University, Nashville, TN  37208, USA}

\author[0009-0000-9360-4759]{Sydney C. Leggio}
\affiliation{Minnesota Institute for Astrophysics, University of Minnesota, Minneapolis, MN 55455, USA}
\affiliation{School of Physics and Astronomy, University of Minnesota, Minneapolis, MN 55455, USA}

\author[0000-0002-9108-5059]{R. Weizmann Kiendrebeogo}
\affiliation{School of Physics and Astronomy, University of Minnesota, Minneapolis, MN 55455, USA}
\affiliation{Laboratoire de Physique et de Chimie de l'Environnement, Universit\'{e} Joseph KI-ZERBO, Ouagadougou, Burkina Faso}
\affiliation{Artemis, Observatoire de la C\^{o}te d'Azur, Universit\'{e} Côte d'Azur, Boulevard de l'Observatoire, F-06304 Nice, France}  

\author[0000-0002-8262-2924]{Michael W. Coughlin}
\affiliation{Minnesota Institute for Astrophysics, University of Minnesota, Minneapolis, MN 55455, USA}
\affiliation{School of Physics and Astronomy, University of Minnesota, Minneapolis, MN 55455, USA}

\author[0000-0001-5857-5622]{Hannah P. Earnshaw}
\affiliation{Division of Physics, Mathematics, and Astronomy, California Institute of Technology, Pasadena, CA 91125, USA}

\author[0000-0003-3703-5154]{Suvi Gezari}
\affiliation{Department of Physics and Astronomy, Johns Hopkins University, 3400 N. Charles Street, Baltimore, MD 21218, USA}
\affiliation{Space Telescope Science Institute, 3700 San Martin Drive, Baltimore, MD 21218, USA}

\author[0000-0002-1984-2932]{Brian W. Grefenstette}
\affiliation{Division of Physics, Mathematics, and Astronomy, California Institute of Technology, Pasadena, CA 91125, USA}

\author[0000-0003-2992-8024]{Fiona A. Harrison}
\affiliation{Division of Physics, Mathematics, and Astronomy, California Institute of Technology, Pasadena, CA 91125, USA}

\author[0000-0002-5619-4938]{Mansi M. Kasliwal}
\affiliation{Division of Physics, Mathematics, and Astronomy, California Institute of Technology, Pasadena, CA 91125, USA}

\author[0000-0003-2528-3409]{Brett M. Morris}
\affiliation{Space Telescope Science Institute, 3700 San Martin Drive, Baltimore, MD 21218, USA}

\author[0000-0002-9599-310X]{Erik Tollerud}
\affiliation{Space Telescope Science Institute, 3700 San Martin Drive, Baltimore, MD 21218, USA}

\author[0000-0003-1673-970X]{S. Bradley Cenko}
\affiliation{Astroparticle Physics Laboratory, NASA Goddard Space Flight Center, Greenbelt, MD 20771, USA}
\affiliation{Joint Space-Science Institute, University of Maryland, College Park, MD 20742, USA}

\begin{abstract}
\Ac{UVEX} is a wide-field ultraviolet space telescope selected as a NASA \ac{MIDEX} mission for launch in 2030. \ac{UVEX} will undertake deep, cadenced surveys of the entire sky to probe low mass galaxies and explore the \ac{UV} time-domain sky, and it will carry the first rapidly deployable \ac{UV} spectroscopic capability for a broad range of science applications. One of \ac{UVEX}'s prime objectives is to follow up \ac{GW} binary neutron star mergers as \acp{ToO}, rapidly scanning across their localization regions to search for their \ac{KN} counterparts. Early-time multiband ultraviolet light curves of \acp{KN} are key to explaining the interplay between jet and ejecta in binary neutron star mergers. Owing to high Galactic extinction in the ultraviolet and the variation of \ac{GW} distance estimates over the sky, the sensitivity to kilonovae can vary significantly across the \ac{GW} localization and even across the footprint of a single image given \ac{UVEX}'s large \acl{FOV}. Good \ac{ToO} observing strategies to trade off between area and depth are neither simple nor obvious. We present an optimal strategy for \ac{GW} follow-up with \ac{UVEX} in which exposure time is adjusted dynamically for each field individually to maximize the overall probability of detection. We model the scheduling problem using the expressive and powerful mathematical framework of \ac{MILP}, and employ a state-of-the-art \ac{MILP} solver to automatically generate observing plan timelines that achieve high probabilities of kilonova detection. We have implemented this strategy in an open-source astronomical scheduling software package called \acl{M4OPT}, on GitHub at \url{https://github.com/m4opt/m4opt}.
\end{abstract}


\acresetall

\section{Introduction} \label{sec:intro}

In 2017, \ac{LIGO} and Virgo \citep{2017PhRvL.119p1101A} detected a long-duration \ac{GW} inspiral signal, GW170817, at the same time that Fermi \citep{2017ApJ...848L..14G} and INTEGRAL \citep{2017ApJ...848L..15S} recorded a short \ac{GRB}, GRB~170817A. The alert sprang traps that had been set by hundreds of telescopes worldwide \citep{2014ApJS..211....7A,2016ApJ...826L..13A} which quickly found the optical counterpart \citep{2017Sci...358.1556C}, AT2017gfo. As a measure of the degree to which the event focused the efforts of astronomers everywhere, the author list of \citet{2017ApJ...848L..12A} runs to 24 pages!

The scientific harvest from this one event was remarkable. It fulfilled a three-decade-old dream of using \acp{GW} as ``standard sirens'' to measure the Hubble constant \citep{1986Natur.323..310S,2017Natur.551...85A}. Moreover, it proved once and for all the hypotheses that \ac{NS} mergers are the central engines of short \acp{GRB} \citep{2013ApJ...776...18F,2017ApJ...848L..13A} and the main cosmic factories of heavy $r$-process elements \citep{1999ApJ...525L.121F,2017Sci...358.1570D,2017Natur.551...80K,2017Sci...358.1583K}.

It had long been understood that such mergers would tidally disrupt their \acp{NS}, and that radioactive decay of the heavy elements synthesized in their hot neutron-rich ejecta would fuel transients \citep{1974ApJ...192L.145L,1989Natur.340..126E,1998ApJ...507L..59L} that came to be called \acp{KN}.

In those early days, astronomers assumed that the ejecta would have opacities similar to those in \acp{SN} and predicted fairly bright \ac{KN} light curves that peaked in the optical or \ac{UV} and that would be fairly easy to detect. However, further study of the atomic structure of lanthanides led to the realization that their dense absorption spectra would lead to line-blanketing in the optical, containing the radiation and only letting it leak out much more slowly and at longer wavelengths, in the infrared \citep{2013ApJ...774...25K}. Observers grimly realized that although \acp{KN} were still among the most promising counterparts of \ac{NS} mergers \citep{2012ApJ...746...48M}, they would be much dimmer, redder, and harder to detect than previously expected. Just how dim and red might depend sensitively on the masses and spins of the compact objects and whether either of them is a black hole (for a review, see \citealt{2020LRR....23....1M}).

Although these later and more sober predictions agreed remarkably well with the observed spectral sequence of AT2017gfo at times later than a few days \citep{2017Sci...358.1570D,2017Natur.551...80K,2017Sci...358.1583K,2017Natur.551...67P,2017ApJ...851L..21V}, contrary to those expectations it was quite blue and featureless at the earliest observed times, less than a day after the merger \citep{2017Sci...358.1574S}.

In the fallout from GW170817, the cause of this early optical and \ac{UV} emission remains one of the most enduring mysteries. The blue emission could be radioactively powered but result from a geometrically distinct outflow component with higher velocity and\,/\,or lower lanthanide fraction \citep{2017ApJ...848L..18N} or could result from the shock caused by the ``cocoon'' interaction between the ejecta and the nascent jet \citep{2017Sci...358.1559K,2018MNRAS.479..588G,2018ApJ...855..103P}. Lacking for GW170817, early-time \ac{UV} observations, less than 12~hr after merger, could handily settle the debate \citep{2018ApJ...855L..23A}.

\subsection{The Coming UV Time Domain Revolution}

More generally, there is a recognized gap in transient discovery capability in the \ac{UV}, and an acknowledged need for a space-based \ac{UV} wide-field time-domain survey \citep{2014AJ....147...79S}. To meet this need, some of the authors proposed Dorado (n\'{e}e GUCI, \citealt{2019AAS...23421203C,2023ApJ...944..126D}) to NASA as a \ac{MoO}.

\begin{figure}
    \includegraphics[width=\columnwidth]{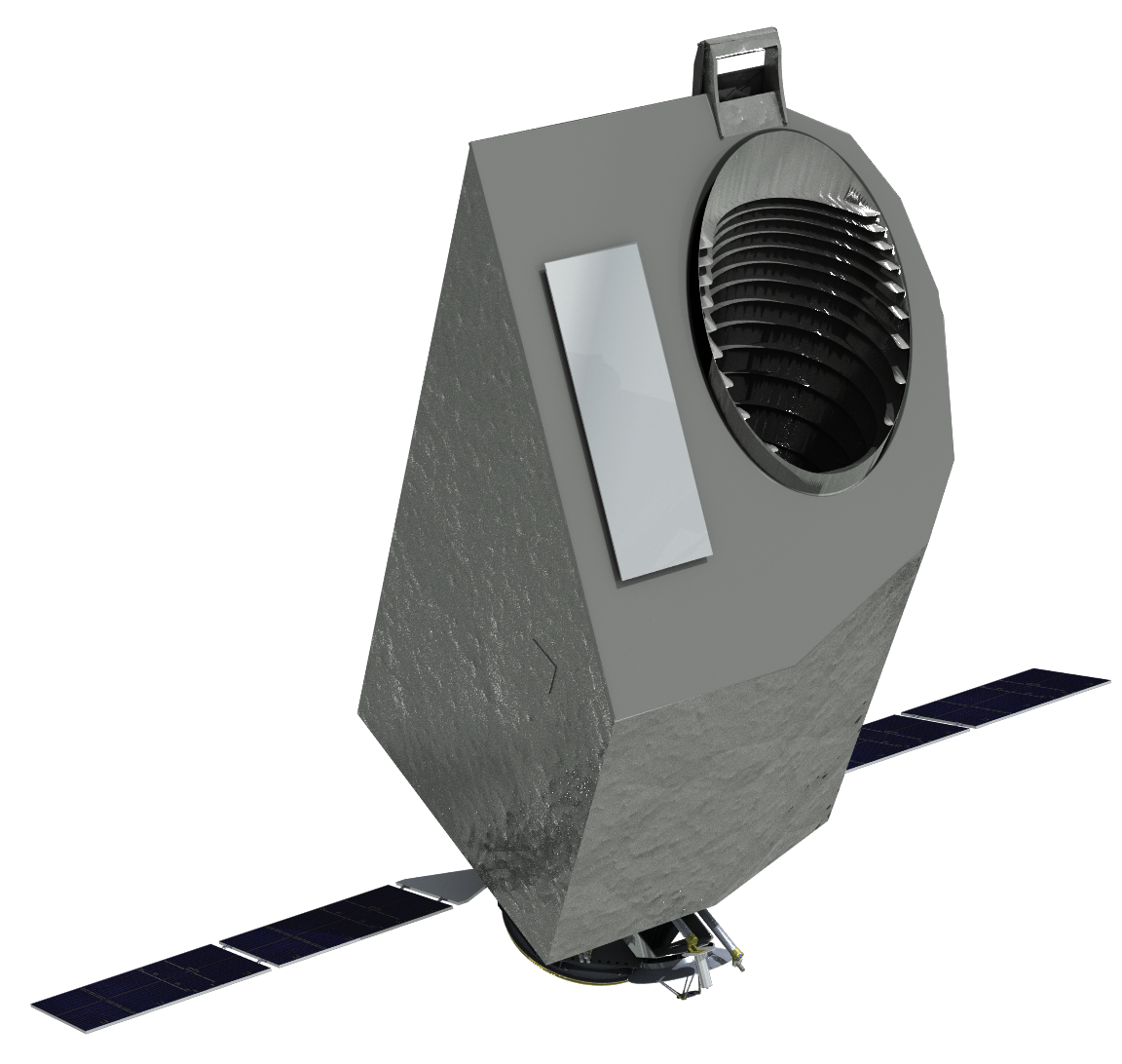}
    \caption{\label{fig:render}A rendering of \ac{UVEX}. Reproduced from the \acf{CSR}.}
\end{figure}

Although NASA made no \ac{MoO} selection in that cycle, shortly thereafter the \citet{2021pdaa.book.....N} recommended in the 2020 decadal survey that ``NASA should establish a time-domain program to realize and sustain the necessary suite of space-based electromagnetic capabilities required to study transient and time-variable phenomena, and to follow up multi-messenger events.'' NASA soon selected a much larger and more capable mission called \acl{UVEX} (\acsu{UVEX}; \citealt{2021arXiv211115608K}; see rendering in Fig.~\ref{fig:render})\footnote{\url{https://www.uvex.caltech.edu}} as the next \ac{MIDEX}.

\begin{figure}
    \includegraphics[width=\columnwidth]{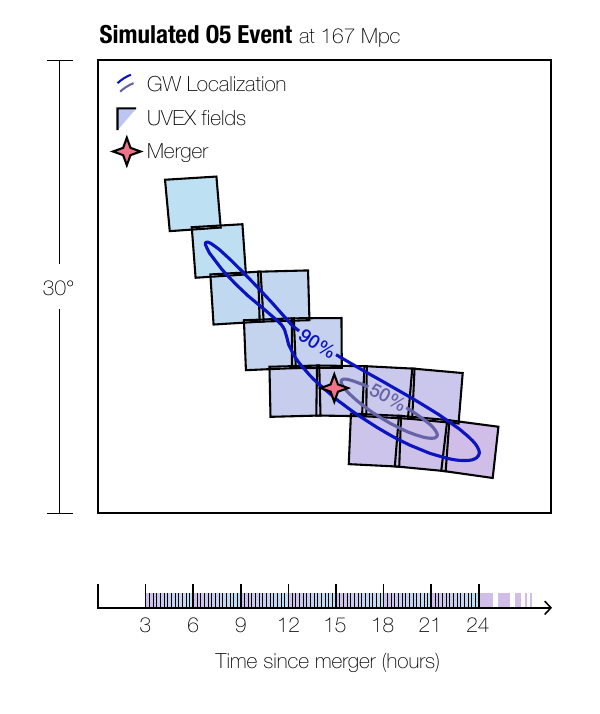}
    \caption{\label{fig:uvex-tiling}An example of a \ac{ToO} observation sequence with \ac{UVEX} to follow up a \ac{GW} event. Adapted from Fig.~D-8 from the \ac{UVEX} \ac{CSR} submitted to NASA \ac{HQ}. This image was also reproduced as Fig.~14 in \citet{2021arXiv211115608K}.}
\end{figure}

\ac{UVEX} will have a wide, $3.5\arcdeg \times 3.5\arcdeg$ \ac{FOV} camera that will take images simultaneously in both a \acl{NUV} (\acsu{NUV}; 1390--1900~\AA) and a \acl{FUV} (\acsu{FUV}; 2030--2700~\AA) band reaching typical depths of $>$24.5~mag in a typical 900~s dwell. The imaging \ac{PSF} diameter of about $\sim$2$\arcsec$ will be well-matched to ground-based follow-up. It will also have a long-slit spectrograph. \ac{UVEX} will perform a mix of surveys with different sky coverage and cadence, and will observe the entire sky to at least $>$25.8~mag over the duration of the prime mission. \ac{UVEX} will not just perform a transformative all-sky time-domain \ac{UV} survey. It will also be able to perform \acp{ToO} to follow up \ac{GW} mergers and other multimessenger phenomena (see Fig.~\ref{fig:uvex-tiling}). With its wide \ac{FOV} and two \ac{UV} bandpasses, it should begin to probe the physical mechanism of early-time emission in \acp{KN}.

Meanwhile, other space agencies are working on complementary ultraviolet facilities. For example, Israel has a mission called \acl{ULTRASAT} (\acsu{ULTRASAT}; \citealt{2024ApJ...964...74S}) for which NASA has agreed to provide the launch. \ac{ULTRASAT} will have an even larger \ac{FOV} ($14\arcdeg \times 14 \arcdeg$) than \ac{UVEX} and will be ready for flight a few years before, but with some tradeoffs: it is much less sensitive (22.5~mag in 900~s), has a much larger \ac{PSF} ($8 \farcs 3$ diameter), only a single imaging band (2300--2900~\AA), and no spectrograph. Still, \ac{UVEX} and \ac{ULTRASAT} together will make a potent discovery engine for \ac{UV} transients in general and \ac{GW} counterparts in particular.

\subsection{A Multi-mission Multi-messenger Observation Planning Toolkit}

NASA selects Explorer-class missions on the basis of a \acf{CSR} which undergoes several reviews, culminating in an in-person \ac{SV}. In \citet{2025PASP..137e4101C}, we elaborated upon the case for \ac{GW} follow-up with \ac{UVEX} as we presented it to NASA in the \ac{CSR}. That \ac{UVEX} study leveraged the same \ac{GW} \ac{ToO} analysis and strategy that we had developed for the Dorado \ac{CSR}. We envisioned our observation planning software, \texttt{dorado-scheduling},\footnote{\url{https://github.com/nasa/dorado-scheduling}} as an early draft of what would eventually evolve into the observatory's real-time science operations software and a part of its public guest observer toolkit. Therefore, both studies benefited from an unusually high level of fidelity and realism for so early in their respective mission lifecycles.

The \texttt{dorado-scheduling} package considers, for any given \ac{ToO}, the time-varying \ac{FOR} constraints of the spacecraft, the spacecraft's slew time between any two target fields, the footprint on the sky of the instrument's \ac{FOV}, and the probability distribution of the true but unknown position of the source --- the \ac{GW} localization map. Its purpose is to find the sequence of observations that maximizes the probability of detecting the counterpart. This can be surprisingly challenging: as with similar ``hard'' optimization problems (e.g. the ``traveling salesman'' problem), the combinatorial scale makes a brute-force search of all possible observation sequences out of the question. On the other hand, simple strategies can yield unacceptable results. For example, observing the fields in descending order of probability may be a poor strategy if it causes one to miss fields that pass out of the \ac{FOR} early on, or if it results in long slew times. Moreover, it may be labor-intensive to adjust or rewrite any handmade strategy to re-tune it for changes to the mission design. Instead of such sub-optimal heurstics, \texttt{dorado-scheduling} directly and globally optimizes the predicted probability of detection using the versatile and expressive mathematical framework of \acl{MILP} (\acsu{MILP}; see textbooks of \citealt{9781118166000} and \citealt{williams2013}), and employs IBM's state-of-the-art \ac{MILP} solver, CPLEX,\footnote{\url{https://www.ibm.com/products/ilog-cplex-optimization-studio}} to search for the global optimum.

Responding to queries during the Dorado \ac{SV} about varying extinction and foregrounds, we added a powerful capability to dynamically optimize the exposure time of each field to adjust for spatial variations in sensitivity. This refinement brought in several new possible tradeoffs to increase detection probability. For example, the \ac{GW} localization is a distribution over both sky location and distance, and there is a Malmquist-like tendency for regions of high probability on the sky to favor large distances \citep{2016ApJ...829L..15S}, so at times it is advantageous to observe lower probability fields that require less exposure time. Furthermore, \ac{UV} sensitivity varies significantly across the sky. For instance, it may be beneficial to spend less time observing high-probability fields that also have high Milky Way dust extinction. The \ac{MILP} framework allowed us to take advantage of these additional degrees of freedom without needing to meticuluously craft the observing strategy itself.

In our simulations, adaptive exposure time greatly improved the detection efficiency of the mission. In earlier related works, \citet{2016ExA....42..165C} derived power-law expressions for optimal exposure time under simplified conditions. \citet{2017ApJ...834...84C} varied exposure times but without considering field-to-field variation in distance, variation in sensitivity, or \ac{FOR} constraints. \citet{2020A&C....3300425H} employed exposure time maps to select feasible fields for \ac{ToO} observations. \citet{2021RAA....21..308L} varied exposure time in accordance with the expected time-dependence of the light curve. However, we believe that our adaptive optimization of exposure time for each field, accounting for spatial variation in distance estimate and sensitivity, to globally maximize the probability of detection, is novel. Unfortunately, this code was not used in the \ac{UVEX} study or in \citet{2025PASP..137e4101C}.

The \texttt{dorado-scheduling} code also had the major limitation that it was released under the obscure and anachronistic \ac{NOSA}, which placed severe obstacles to our own collaborators using or contributing to it. The lawyers at \ac{GSFC} continue to require all \ac{GSFC} scientists to employ \ac{NOSA} even though it has has been rejected by the \citet{FSF}, the \citet{NAP25217}, and even the \citet{SMD}. Fortunately, NASA \ac{HQ} intervened in this case and we were permitted to establish a new, truly open-source, permissively licensed software project, which anyone can contribute to and use. However, this required a rewrite which took several years.

In this paper, we finally present \ac{M4OPT}, which has many other advances over earlier work:
\begin{itemize}
    \item It is released under a permissive, mainstream license (the Berkeley Software Distribution, or BSD, license) to promote adoption and contribution by the community.
    \item It is designed from the start to support multiple missions, including \ac{UVEX} and \ac{ULTRASAT}. We envision supporting both space- and ground-based observatories in the future.
    \item It can adjust exposure times given the anticipated absolute magnitude and the three-dimensional \ac{GW} localization distribution in sky position and distance.
    \item It can dynamically vary the exposure time of each field to pierce through spatial variations in foregrounds (zodiacal light, Galactic diffuse emission) and dust extinction.
    \item It can be given the anticipated absolute magnitude of the source as either a point estimate or a distribution with Gaussian uncertainty.
    \item The dynamic exposure times are made possible by a \texttt{numpy} \citep{harris2020array} vectorized \ac{ETC} which enables large parameter sweeps of synthetic photometry calculations which are otherwise prohibitively slow with synphot \citep{2018ascl.soft11001S} alone.
    \item It models additional spacecraft dynamics effects, including the roll angle of the telescope which is determined by solar power requirements.
    \item It already complies fully with \ac{NPR} 7150 \citep{NPR7150} software engineering practices for ``Class C'' software, and has 95\% test coverage.
    \item It is deeply integrated with the Astropy \citep{2013A&A...558A..33A,2018AJ....156..123A} ecosystem, and has an interface that is based on Astropy coordinates, units, and model classes.
\end{itemize}

\begin{figure*}
    \begin{interactive}{animation}{and/example.gif}
        \includegraphics[width=0.9\textwidth]{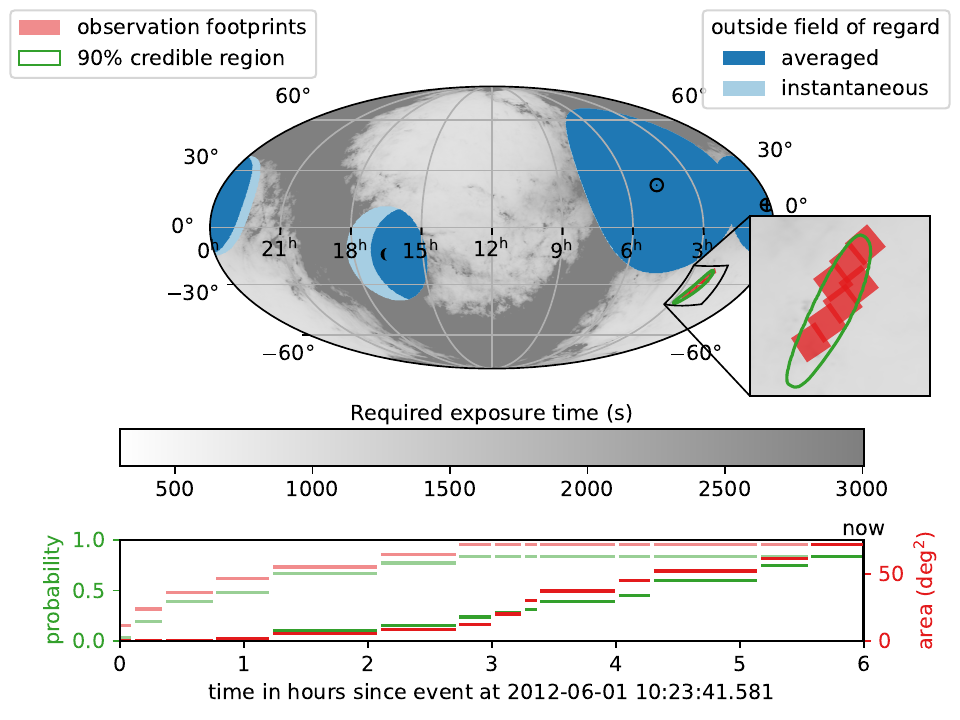}
    \end{interactive}
    \caption{\label{fig:example}An example \ac{M4OPT} observing plan. Top panel and inset: The footprints of the observations filled in pink, the 90\% credible region of the \ac{GW} localization outlined in green, the region that is always outside of the \ac{FOR} (referred to in the legend as the ``averaged'' \ac{FOR}) filled in deep blue, and the region that is outside of the instantaneous \ac{FOR} filled in light blue. Bottom panel: The timeline of the accumulation of detection probability and area. The region that has been covered by one visit is shown in pale red and green, whereas the region that has been covered by two visits is shown in deep red and green. This figure is available as an animation.}
\end{figure*}

In this paper, we describe the mathematical approach of \ac{M4OPT}, and then use it to produce realistic \ac{ToO} observing sequences with \ac{UVEX} for simulated \ac{GW} events. An example \ac{M4OPT} observing plan is shown in Fig.~\ref{fig:example} and Table~\ref{table:example}. We show that the new dynamic exposure time capabilities dramatically increases the probability of \ac{KN} detection.

\begin{deluxetable*}{lrcccccccccccrc}
    \tablecaption{\label{table:example}An Example \ac{M4OPT} Observing Plan (same as Fig.~\ref{fig:example})}
    \tablehead{
        &&& \multicolumn3c{Location\tablenotemark{\scriptsize{a}}}
        & \multicolumn3c{Target\tablenotemark{\scriptsize{b}}}
        & \multicolumn{2}{c}{Sky map}
        \\
        \colhead{Time\tablenotemark{\scriptsize{c}}} &
        \colhead{Dur.\tablenotemark{\scriptsize{d}}} &
        \colhead{Action} &
        \colhead{$x$} &
        \colhead{$y$} &
        \colhead{$z$} &
        \colhead{R.A.} &
        \colhead{Decl.} &
        \colhead{Roll} &
        \colhead{Prob.\tablenotemark{\scriptsize{e}}} &
        \colhead{Dist.\tablenotemark{\scriptsize{f}}} &
        \colhead{Lim.\tablenotemark{\scriptsize{g}}} &
        \colhead{Dust\tablenotemark{\scriptsize{h}}} &
        \colhead{Sky\tablenotemark{\scriptsize{i}}} &
        \colhead{Det. Prob.\tablenotemark{\scriptsize{j}}}
    }
    \startdata
    \input{tables/example.tex}
    \enddata
    \tablenotetext{a}{Cartesian coordinates (Mm) of the spacecraft in the Earth-fixed (\acl{ITRS}) frame.}
    \tablenotetext{b}{Target location (deg) in the equatorial (\acl{ICRS}) frame.}
    \tablenotetext{c}{UTC start time of scheduled action on the arbitrary date of 2012-06-01.}
    \tablenotetext{d}{Duration (s) of scheduled action.}
    \tablenotetext{e}{Integral of the \ac{GW} probability sky map over this field: probability that the true, but unknown, position of the source is within the footprint.}
    \tablenotetext{f}{A posteriori mean \ac{GW} distance estimate (Mpc) over this field.}
    \tablenotetext{g}{Median limiting magnitude (AB mag) over this field.}
    \tablenotetext{h}{Median dust extinction in band (AB mag) over this field.}
    \tablenotetext{i}{Median sky surface brightness in band (AB mag arcsec$^{-2}$) over this field.}
    \tablenotetext{j}{Contribution by this field to the predicted probability of detection.}

\end{deluxetable*}

\section{MILP Problem Formulation}

\Ac{LP} is a mathematical optimization formalism in which one represents an objective as a linear combination of decision variables, subject to constraints that take the form of a system of linear inequalities. The canonical form a \ac{LP} is
\begin{alignat*}{4}
    \text{Find}\quad && &\mathbf{x} \in \mathbb{R}^N \\
    \text{that maximizes}\quad && \mathbf{c}^\mathsf{T} &\mathbf{x} \\
    \text{subject to}\quad && \mathbf{A} &\mathbf{x} \leq 0 \\
    \text{and}\quad && &\mathbf{x} \geq 0.
\end{alignat*}
\ac{LP} is useful for problems of resource allocation. If instead of being reals, certain of the decision variables are required to be integers, then the problem is called \acf{MILP}. \ac{MILP} allows one to model situations where some decision variables represent alternative courses of action or where some constraints are Boolean in nature. The textbook chapters of \citet{9781118166000.ch3} and \citet{williams2013model} serve as useful translation dictionaries from logical constraints to integer linear ineqalities.

There are a variety of free and open source \ac{MILP} solvers; the most readily available for Python users is HiGHS \citep{huangfu2018parallelizing} which now can be called directly from SciPy \citep{2020NatMe..17..261V}. However, the best commercial \ac{MILP} solvers are much faster and can handle much larger problems than any current open-source software \citep{koch2011miplib,huangfu2018parallelizing}. Although \ac{M4OPT} itself is open source, it calls the commercial \ac{MILP} solver CPLEX by IBM. CPLEX is free for academic and research use by students, faculty, and staff at academic institutions. For non-academic users, its pricing is comparable in cost to mid-range computer-aided design software.

Common applications of \ac{MILP} include planning and scheduling. There are already many noteworthy and successful uses of \ac{MILP} to astronomical observation planning. \ac{LCO} uses it to allocate observations on a global queue-scheduled network of telescopes \citep{2014SPIE.9149E..0ES}. \ac{ZTF} uses it to multiplex several sky surveys with different coverage and cadence requirements on a single telescope \citep{2019PASP..131f8003B}, and our colleagues have proposed using it to schedule \ac{ToO} observations with \ac{ZTF} \citep{2022ApJ...935...87P}. It has been applied to scheduling observing programs on \acl{ALMA} (\acsu{ALMA}; \citealt{2016A&C....15...90S}) and exoplanet searches on Keck \citep{2024AJ....167...33H}.

\subsection{The Maximum Weighted Coverage Problem}

\begin{figure}
    \centering
    \includegraphics[width=0.6\columnwidth]{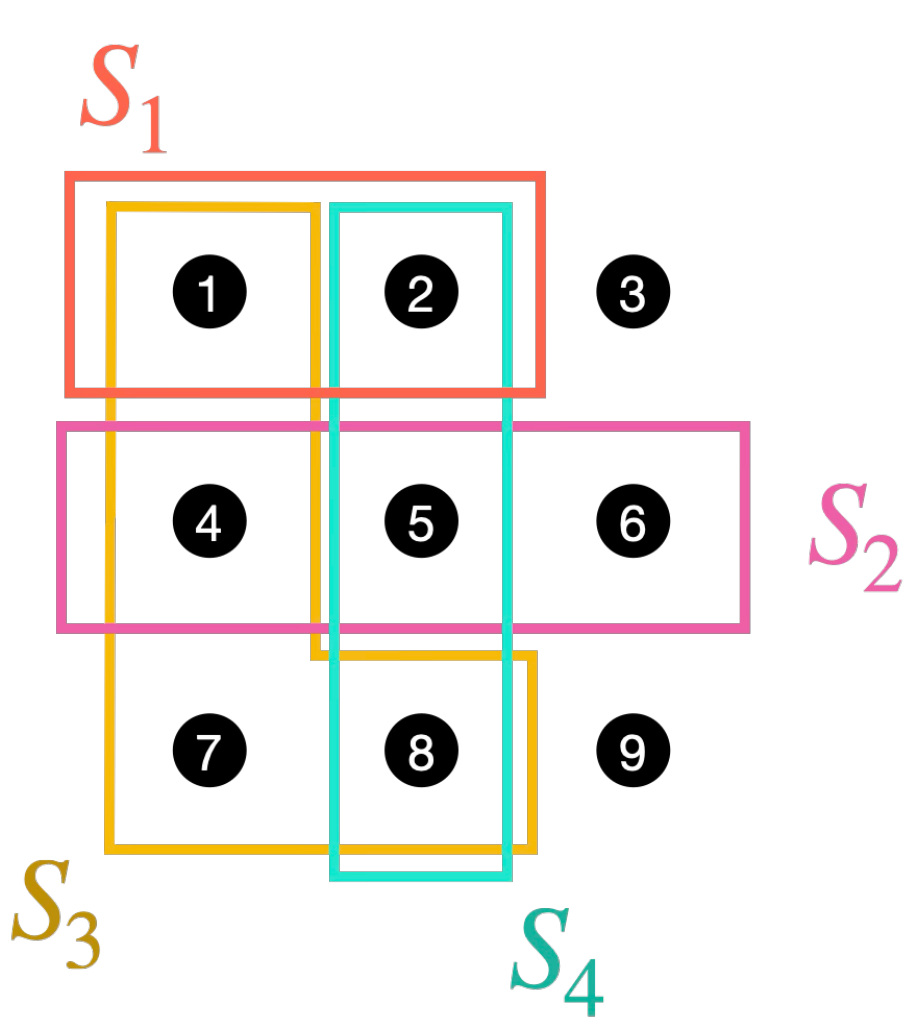}
    \caption{\label{fig:max-weighted-coverage}Illustration of \ac{MWC}. Four sets $S = \{S_1, S_2, S_3, S_4\}$ are represented as regions with colored borders. The elements of those sets are represented by black numbered circles.}
\end{figure}

The classic \ac{MWC} has a \ac{MILP} representation that is at the core of our \ac{ToO} planning problem. In \ac{MWC}, one has a finite sequence of real-valued weights, $(w_j)_j$, and a finite set of sets, $S = \{S_i\}_i$, over the integers, $\forall i: S_i \subset \mathbb{Z}$. The objective is to find a subset $S^\prime \subseteq S$, with a maximum cardinality $|S^\prime| \leq k$, that maximizes the sum over all of the weights $\sum_{j \in \bigcup S^\prime} w_j$. This is illustrated in Fig.~\ref{fig:max-weighted-coverage}. The \ac{MWC} has a straightforward \ac{MILP} representation:
\begin{align*}
\text{Maximize}\quad &\sum_j y_j w_j \\
\text{subject to the constraints}\quad &\sum_i x_i \leq k \\
\text{and}\quad &\sum_{\mathclap{i \mid j \in S_i}} x_i \geq y_j.
\end{align*}

\begin{figure}
    \includegraphics[width=\columnwidth]{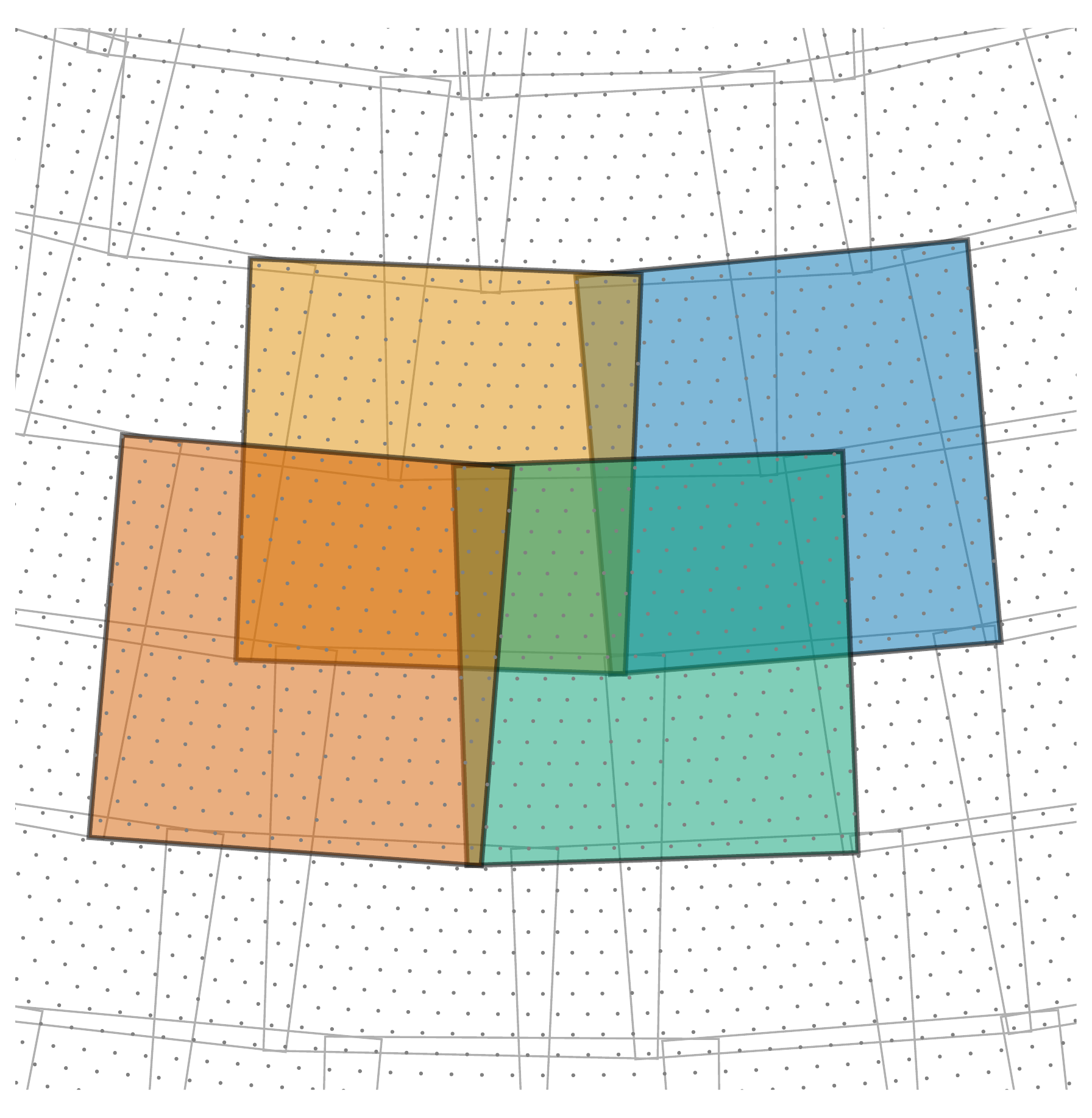}
    \caption{\label{fig:overlapping-fields}\ac{MWC} applied to coverage of a region on the sky by multiple partially overlapping fields. The centers of \ac{HEALPix} pixels are marked with gray dots.}
\end{figure}

The most basic version of our \ac{ToO} problem is that we want to maximize the total integrated probability that the true, but unknown, position of the source is within one or more of the footprints that we select to observe. Each \ac{GW} alert comes with a localization sky map that gives the posterior probability distribution of the sky position as an image that is sampled on a \acl{HEALPix} (\acsu{HEALPix}; \citealt{2005ApJ...622..759G}) grid. The weights $(w_j)_j$ are the probability values and the subsets $(S_i)_i$ are the \ac{HEALPix} pixels contained within the footprints of the \ac{FOV} on a reference grid of allowed pointings of the telescope (see Fig.~\ref{fig:overlapping-fields}). The solution of the \ac{MWC} problem gives us the optimal pointings of the telescope and the \ac{HEALPix} pixels contained within all of the planned observations.

\subsection{Scheduling Constraints}

Scheduling problems in which there are time intervals that must be disjoint are also prominent in classic applications of \ac{MILP}. Consider two intervals, both of duration $\delta$, and each centered on the real-valued event times $t_0$ and $t_1$ respectively, and both intervals of duration $\delta$. Suppose also that the event times $t_0$ and $t_1$ are both restricted to the range from 0 to $M$. If the two intervals must be disjoint, then we can express the situation with the following system of ineqalities:
\begin{align*}
    0 \leq t_0 \leq M \\
    0 \leq t_1 \leq M \\
    \left| t_1 - t_0 \right| \geq \delta.
\end{align*}
The last inequality is nonlinear because it contains the absolute value function. By introducing a binary decision variable $a$, we can transform it into two linear inequalities,
\begin{align*}
    t_1 - t_0 - \delta &\leq M (1 - a) \\
    t_0 - t_1 - \delta &\leq M a.
\end{align*}

In our \ac{ToO} problem, the times $t_0$ and $t_1$ represent the times of observations of two different fields that must be executed by a single telescope, and the duration $\delta$ represents the exposure time plus slew time. By extending to more than two intervals we can model scheduling of a sequence of observations by one telescope.

\subsection{The Full Problem}

The full \ac{ToO} problem as it is modeled and solved by \ac{M4OPT} has elements of the \ac{MWC} and elements of scheduling. The reader may find the full \ac{MILP} problem setup in Appendix~\ref{sec:details}, in which we develop three successively more elaborate models, finally arriving at the full problem as it is implemented by \ac{M4OPT}. These three stages are:

\paragraph{Section~\ref{sec:fixed-exptime}, Fixed Exposure Time} We may observe any fields within the instantaneous \ac{FOR}. Each field must be visited twice with a minimum cadence between visits. Every field has the same fixed exposure time. The objective is to maximize the probability that the true (but unknown) position of the source is within any of the selected fields. The probability of detection is integrated over the two-dimensional \ac{GW} localization map.

\paragraph{Section~\ref{sec:variable-exptime}, Variable Exposure Time} As in Section~\ref{sec:fixed-exptime}, but each field's exposure time may vary. We assume that the apparent magnitude of the source is fixed with respect to time and known precisely, and the objective is to maximize the probability that the limiting magnitude at the true (but unknown) location of the source exceeds that apparent magnitude. The probability of detection is integrated over the two-dimensional \ac{GW} localization map and employs a sky map of the limiting magnitude as a function of exposure time.

\paragraph{Section~\ref{sec:absmag-distn}, Variable Exposure Time with Prior Distribution of Absolute Magnitude} As in Section~\ref{sec:variable-exptime}, but we no longer know the apparent magnitude of the source. Instead we have a luminosity function, a probability distribution over its true (but unknown) absolute magnitude, which is fixed with respect to time. We also have a three-dimenisional \ac{GW} localization map in right ascension, declination, and luminosity distance. The probability of detection is integrated over the 3D localization map and the luminosity function.

\section{Exposure Time Estimation}
\label{sec:etc}

\begin{figure}
    \includegraphics[width=\columnwidth]{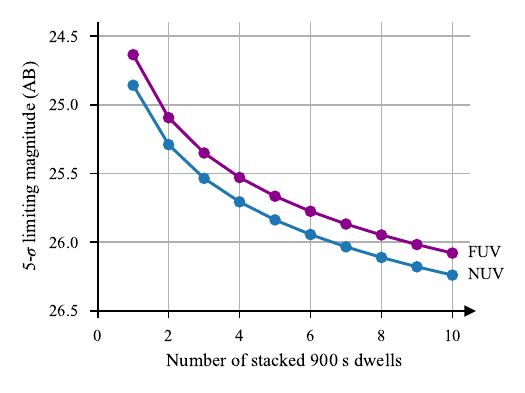}
    \caption{\label{fig:etc}Median limiting magnitude, averaged over target coordinates and observation time.}
\end{figure}

The reference \ac{ETC} for \ac{UVEX} is still closed source because the instrument is still evolving in small details. We use \ac{M4OPT}'s open-source toy model of the instrument performance that roughly reproduces the public sensitivity curve plot on the \ac{UVEX} web site\footnote{\url{https://www.uvex.caltech.edu/page/for-astronomers}; accessed 2025 January 26.} (see Fig.~\ref{fig:etc}). The parameters of the toy model are listed in Table~\ref{tab:etc}.

\begin{deluxetable}{lc}
    \tablecaption{\label{tab:etc}\ac{ETC} Toy Model Parameters}
    \tablehead{
        \colhead{Quantity} & \colhead{Value}
    }
    \startdata
    Aperture diameter & 75 cm\\
    Pixel scale & 1$\arcsec$ pix$^{-1}$ \\
    \ac{PSF} sharpness\tablenotemark{\scriptsize{a}} & $1/(4 \pi)$ \\
    Gain & 0.85 \\
    Read noise & 2 ct \\
    Dark noise & $10^{-3}$ ct s$^{-1}$ \\
    \ac{NUV} response\tablenotemark{\scriptsize{b}} & $0.2 \sqrt{2 \pi} \phi[(\lambda - 2300 \text{ \AA}) / (180 \text{ \AA})]$ \\
    \ac{FUV} response\tablenotemark{\scriptsize{b}} & $0.15 \sqrt{2 \pi} \phi[(\lambda - 1600 \text{ \AA}) / (100 \text{ \AA})]$
    \enddata
    \tablenotetext{a}{We assume optimal \ac{PSF} photometry with Nyquist pixel sampling, that is, a sharpness of $1/4\pi$ \citep{2005MNRAS.361..861M}.}
    \tablenotetext{b}{The function $\phi(x)$ is the standard normal distribution.}
\end{deluxetable}

We need to perform about 1~million \ac{ETC} calculations at different sky positions and apparent magnitudes for each event. The \texttt{synphot} Python package \citep{2018ascl.soft11001S} is the standard tool for synthetic photometry. Unfortunately, it would be prohibitively time-consuming to set up and evaluate 1~million \texttt{synphot} scenarios due to the overhead of creating all of the Python objects involved. Our \texttt{m4opt.synphot} module accelerates synthetic photometry calculations by using the symbolic algebra package \texttt{sympy} \citep{10.7717/peerj-cs.103} to rearrange the \texttt{synphot} model tree into additive components with separable spatial dependence that can be integrated separately and then added back together. For nonlinear components (i.e., extinction), we do a parameter sweep and employ interpolation. The result is many orders of magnitude faster than \texttt{synphot} and contributes only about a second of run time to the scheduler.

There are three spatially dependent spectral components in the \ac{ETC}:

\paragraph{Zodiacal light}
We modeled zodiacal background light by taking the ``high'' sky background spectrum from the Hubble \ac{STIS} Instrument Handbook \citep{2024stis.rept....5R} and scaling it by a spatially dependent factor resulting from bilinear interpolation of the tables of \citet{1998A&AS..127....1L}. In the future, we plan to replace the spatial dependence with \texttt{zodipy} \citep{2022A&A...666A.107S,2024JOSS....9.6648S}, which has a higher spatial resolution and is valid for locations throughout the Solar System beyond Earth orbit.

\paragraph{Galactic diffuse background}
\citet{2014ApJS..213...32M} provides piecewise cosecant fits to the surface brightness of Galactic diffuse emission in the two GALEX bands. We employ these fits to get the spatial dependence, and obtain the wavelength dependence by interpolating and extrapolating linearly through the two bands.

\paragraph{Dust extinction}
To model extinction due to dust in the Milky Way, we use the \citet{2023ApJ...950...86G} model (see also \citealt{2009ApJ...705.1320G,2019ApJ...886..108F,2021ApJ...916...33G,2022ApJ...930...15D}) as implemented in the \texttt{dust\_extinction} Python package \citep{2024JOSS....9.7023G}. We employ a fixed total-to-selective extinction ratio $R(V) = 3.1$ and obtain $E(B-V)$ reddening values as a function of sky position using the \citet{2016A&A...596A.109P} model as implemented by the \texttt{dustmaps} Python package \citep{2018JOSS....3..695M}.

\section{Case Study: GW Observations with UVEX}
\label{sec:parameters}

Here we explain the setup of \ac{M4OPT} for \ac{UVEX}.

\paragraph{\Ac{GW} localizations}
We started with the same simulated \ac{GW} localizations as \citet{2025PASP..137e4101C}, which covers LIGO, Virgo, and KAGRA's fifth~(O5) and sixth~(O6) observing runs. The data are publicly archived in \cite{r_weizmann_2025_14585837}. These simulated events were generated using the same methodology as \citet{2022ApJ...924...54P} and \citet{2023ApJ...958..158K}, except that the \ac{S/N} threshold for \ac{GW} detection is set to~10. The localizations were generated with the rapid localization engine BAYESTAR \citep{2016PhRvD..93b4013S} and consist of 3D posterior probability distributions of sky location and luminosity distance \citep{2016ApJ...829L..15S,2016ApJS..226...10S}. Like \citet{2025PASP..137e4101C}, we selected only those events for which the rest frame secondary (lighter) compact object mass $m_2$ was $\leq 3 M_\odot$ to pick only events that could plausibly be \ac{BNS} or \ac{NSBH} binaries.

\paragraph{\Ac{KN} absolute magnitude distribution}
\citet{2021arXiv211115608K} considers two alternative \ac{KN} scenarios: nucleosynthesis powered or shock powered. For the former, they employ the semi-analytic $r$-process heating model of \citet{2020ApJ...891..152H}; for the latter, the analytical shock model of \citet{2018ApJ...855..103P}. For both models, Appendix~E.2 of \citet{2021arXiv211115608K} specifies fiducial parameter ranges and the 90\% credible intervals for the peak absolute magnitude in each band. These absolute magnitude ranges are reproduced in the Table~\ref{tab:kn-abs-mag}. \ac{UVEX} observes in both the \ac{NUV} and \ac{FUV} filters simultaneously, and our objective is to detect the source in both bands. Therefore we should plan observations using the fainter of the two models and the fainter of the two bands: the nucleosynthesis-powered model in FUV, with an absolute magnitude range of $[-14.5, -10.2]$. Assuming that this is the 90\% credible interval of a Gaussian distribution, the absolute magnitude has the approximate distribution
\begin{equation}
    \label{eq:absmag-distn}
    M_\mathrm{NUV} \sim \mathcal{N}(-12.4, 1.3).
\end{equation}

\begin{deluxetable}{lcc}
    \tablecaption{\label{tab:kn-abs-mag}Ranges of Peak Absolute Magnitudes of \Acp{KN}}
    \tablehead{
        & \multicolumn2c{Absolute Magnitude Range} \\
        \colhead{Model} & \colhead{NUV} & \colhead{FUV}
    }
    \startdata
    Nucleosynthesis powered & [-15.6, -12.4] & [-14.5, -10.2] \\
    Shock powered & [-17.8, -15.3] & [-17.9, -15.0]
    \enddata
    \tablecomments{Adapted from Appendix E.2 of \citet{2021arXiv211115608K}.}
\end{deluxetable}

\paragraph{Follow-up time window}
\citet{2025PASP..137e4101C} required a single epoch of \ac{UVEX} observations to take 3~hr or less. To match this choice, we configure \ac{M4OPT} to plan two visits of each field with a minimum cadence of 30~minutes between repeated visits, with a total elapsed time limit of 6~hr.

\paragraph{Exposure time limits}
The exposure time is allowed to vary adaptively for each field, with a minimum exposure time of $\epsilon_\mathrm{min} = 300$~s. The minimum exposure time corresponds to a single standard \ac{UVEX} imaging exposure. (A standard survey dwell consists of 3 consecutive stacked 300~s exposures.)

\begin{figure}
    \includegraphics[width=\columnwidth]{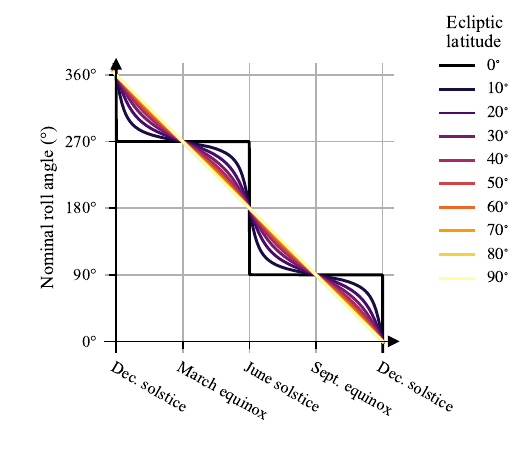}
    \caption{\label{fig:nominal-roll}Nominal roll angle as a function of time for selected ecliptic latitudes and an ecliptic longitude of 270\arcdeg.}
\end{figure}

\paragraph{FOV}
Like most space telescopes, \ac{UVEX} has solar panels that rotate on a \acl{SADA} perpendicular to the telescope boresight (see Fig.~\ref{fig:render}). The position angle of observations is fixed to the nominal roll angle that allows the spacecraft to orient the solar panels perpendicular to the Sun while keeping the cold side of the spacecraft facing away from the Sun. For any given target position, the roll angle goes through one revolution per year. For targets at the ecliptic poles, the nominal roll angle varies linearly with time. For targets in the ecliptic plane, the nominal roll angle flips by 180° twice per year when the target is precisely sunward or anti-sunward (at the solstices if the ecliptic longitude is 90\arcdeg or 270\arcdeg). At all intermediate ecliptic latitudes, the roll angle oscillates smoothly in a manner that interpolates between these extremes (see Fig.~\ref{fig:nominal-roll}). Because the roll angle changes slowly on the timescale of a day, we calculate the nominal roll angle for each field at the time of the event, and leave it fixed at that value for the duration of the observing plan.

\begin{figure}
    \includegraphics[width=\columnwidth]{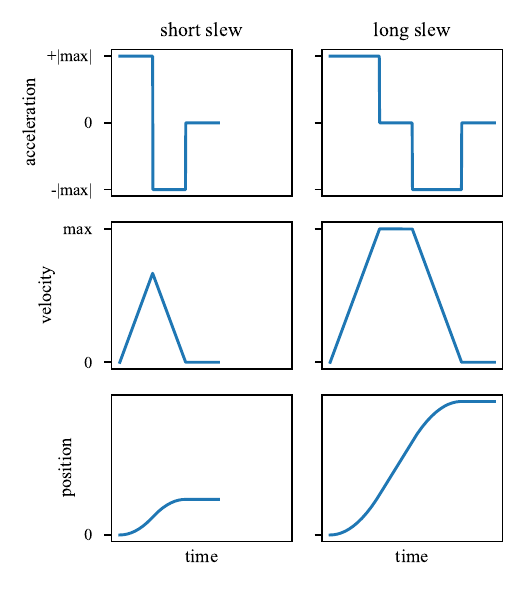}
    \caption{\label{fig:slew}The slew model consists of an acceleration phase at the maximum acceleration, possibly a coasting phase at the maximum angular velocity, a deceleration phase at the maximum acceleration, and a settling phase.}
\end{figure}

\paragraph{Slew time}
To estimate the time to slew between any two fields, we employ a simplified model of the spacecraft dynamics in which the spacecraft has a maximum angular acceleration of $0 \fdg 006$~s$^{-2}$ and maximum angular rate $0 \fdg 6$~s$^{-1}$ about any axis, plus a fixed settling time of 60~s. We assume that the spacecraft executes an eigenaxis maneuver, rotating through the smallest possible angle about a single axis, following a ``bang-bang" angular rate profile illustrated in Fig.~\ref{fig:slew}. (Note that even for a symmetric moment of inertia, this eigenaxis maneuver is \emph{not} the fastest possible slew; see \citealt{1993JGCD...16..446B}.) With this model, it takes 108~s to slew by the width of the \ac{FOV} and 460~s to slew by 180$\arcdeg$.

\begin{figure}
    \includegraphics[width=\columnwidth]{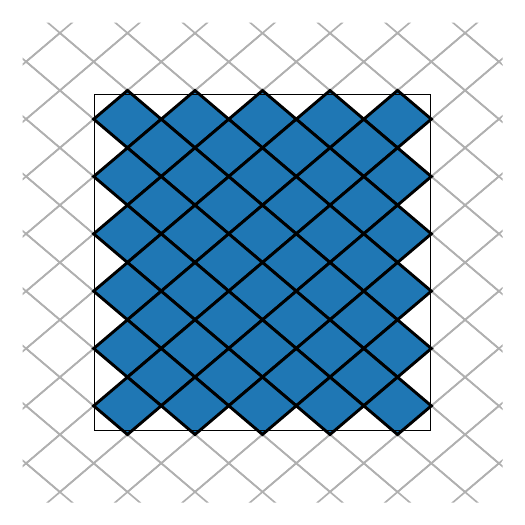}
    \caption{\label{fig:fov}The footprint of the \ac{UVEX} \ac{FOV} discretized on a \ac{HEALPix} grid with $n_\mathrm{side} = 128$. The footprint, or the region on the sky that is contained within the \ac{FOV} in one observation, is shown as the solid gray square. The edges of \ac{HEALPix} pixels are depicted as light gray lines. The HEALPix pixels whose centers are within the footprint are filled in blue.}
\end{figure}

\paragraph{\ac{HEALPix} resolution}
We discretize the footprint of the \ac{FOV} on a \ac{HEALPix} grid with $n_\mathrm{side} = 128$, which is sufficient to resolve the square shape of the \ac{FOV} (see Fig.~\ref{fig:fov}.

\begin{figure}
    \begin{interactive}{animation}{and/skygrid.gif}
        \plotone{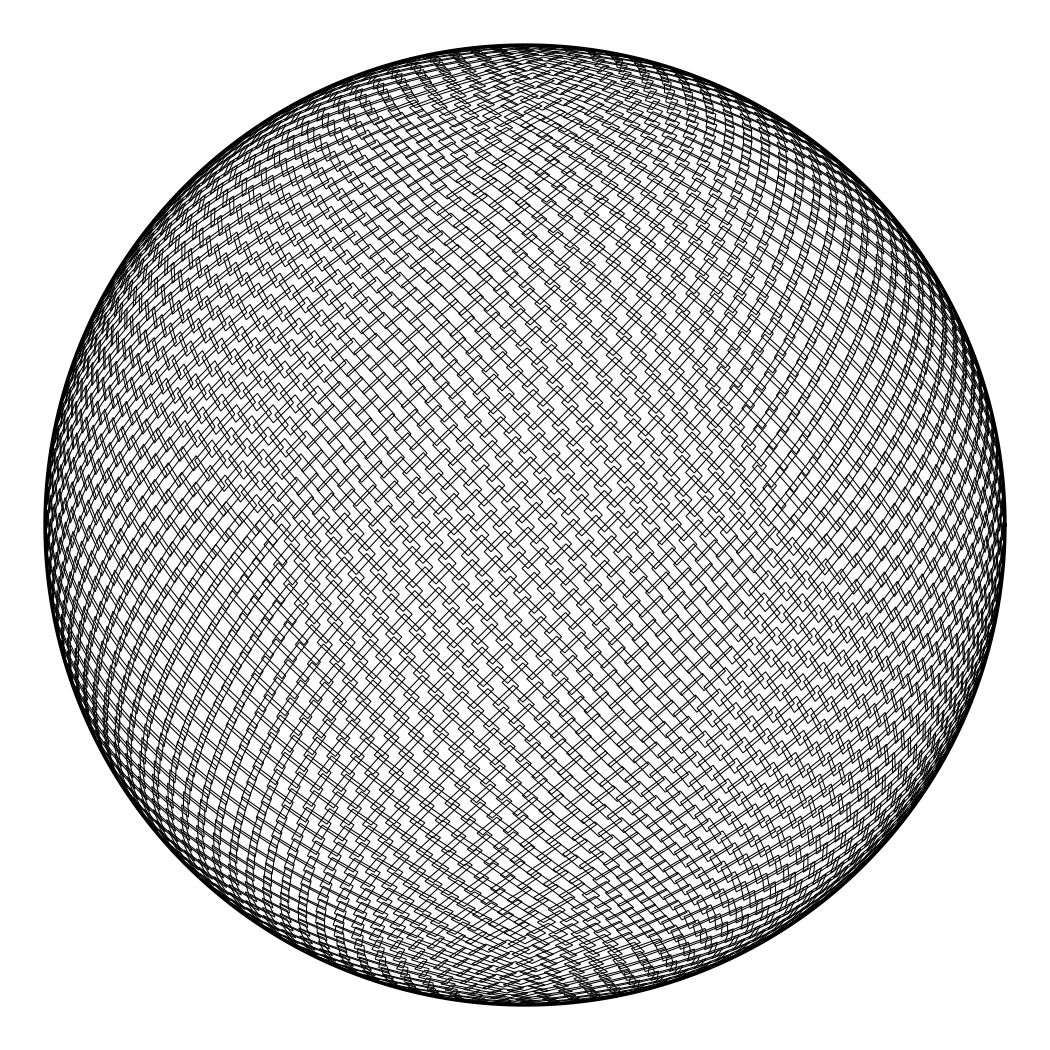}
    \end{interactive}
    \caption{\label{fig:skygrid}Footprints of the \ac{UVEX} \ac{FOV} on the geodesic reference grid. This figure is available as an animation that shows the variation in position angle of each field with time.}
\end{figure}

\paragraph{Field grid}
The centers of the fields are the vertices of a $\{3,5+\}_{21,4}$ icosahedral geodesic polyhedron (see Fig.~\ref{fig:skygrid}) generated using the \texttt{antiprism} software by Adrian Rossiter.\footnote{\url{https://github.com/antiprism/antiprism}} This grid ensures that all of the fields cover the sky without gaps, regardless of roll angle.

\paragraph{Run duration}
As in \citet{2025PASP..137e4101C}, we assumed 1.5~yr of overlap between the \ac{UVEX} prime mission and the \ac{GW} observing run.

\paragraph{Follow-up selection criterion}
We ran \ac{M4OPT} on all simulated events. We consider an event selected for follow-up if the scheduler's objective value $P$ is less than $P^* = 0.1$. As the \ac{MILP} solver runs, it keeps track of both the best objective value of any solution found so far---the ``incumbent'' solution---and an upper bound on the objective value of any solution, including as-yet unexplored solutions. Normally the solver would terminate when the gap between the incumbent objective value and the best bound close to within some numerical tolerance, but we also configure the solver to stop early if the best bound ever drops below $P^*$ because the event in question will not be selected for follow-up.

Recall from Section~\ref{sec:absmag-distn} that the scheduler's objective value is (a numerical approximation of) the probability of detection---integrated over the absolute magnitude, sky position, and distance, none of which are known to the scheduler. We stress that our strategy for selecting which events to trigger is to simply run the scheduler for every single event and proceed if the predicted detection probability is at least 10\%.

This is conceptually very different from the selection criteria in \citet{2025PASP..137e4101C} which is based on the 90\% credible area, $A_\mathrm{90\%}$, and luminosity distance $d_\mathrm{L}$ of the event. However, we can predict analytical thresholds on both of these quantities that will be roughly equivalent to our strategy:
\begin{equation}
    d_\mathrm{L} < d_\mathrm{L}^* = 10^{\frac{1}{5}(x^* - \mu_X + \sigma_X \Phi^{-1}(1-P^*) - 25)}\,\mathrm{Mpc} \label{eq:threshold-distance}
\end{equation}
\begin{equation}
    A_Q < A_Q^* = \left(\frac{\Psi^{-1}(Q)}{\Psi^{-1}(P^*)}\right)^2 \left(\frac{\delta - \beta}{\epsilon_\mathrm{min} n_K}\right)A_\mathrm{FOV} \label{eq:threshold-area}
\end{equation}
\begin{equation}
    \frac{A_Q}{A_\mathrm{FOV}} < \left(\frac{d_\mathrm{L}}{d_\mathrm{L}^*}\right)^{-4} \label{eq:threshold-area-distance}
\end{equation}
$A_Q$ is the $Q$th percentile credible region, $x^*$ is the faintest limiting magnitude at any point on the sky, $A_\mathrm{FOV}$ is the area of the \ac{FOV}, $\Phi^{-1}(x)$ is the inverse of the \ac{CDF} of the standard normal distribution, and $\Psi^{-1}(x) = -2\ln(1 - x)$ is the inverse of the \ac{CDF} of a $\chi^2$ distribution with 2 degrees of freedom.

\subsection{Results}

All simulated events are listed in the online version of Table~\ref{tab:events}. Because these are simulated events, we can determine the probability of detection for each event given its true sky location and distance. To calculate the detection probability, we calculate the limiting absolute magnitude at that true position and distance for the longest exposure that contains that position (or an exposure time of zero if the true sky position is not contained in any planned observation). The detection probability is simply the \ac{CDF} of the absolute magnitude distribution, Eq.~(\ref{eq:absmag-distn}), evaluated at that limiting absolute magnitude.

\begin{figure*}
    \centering
    \gridline{\fig{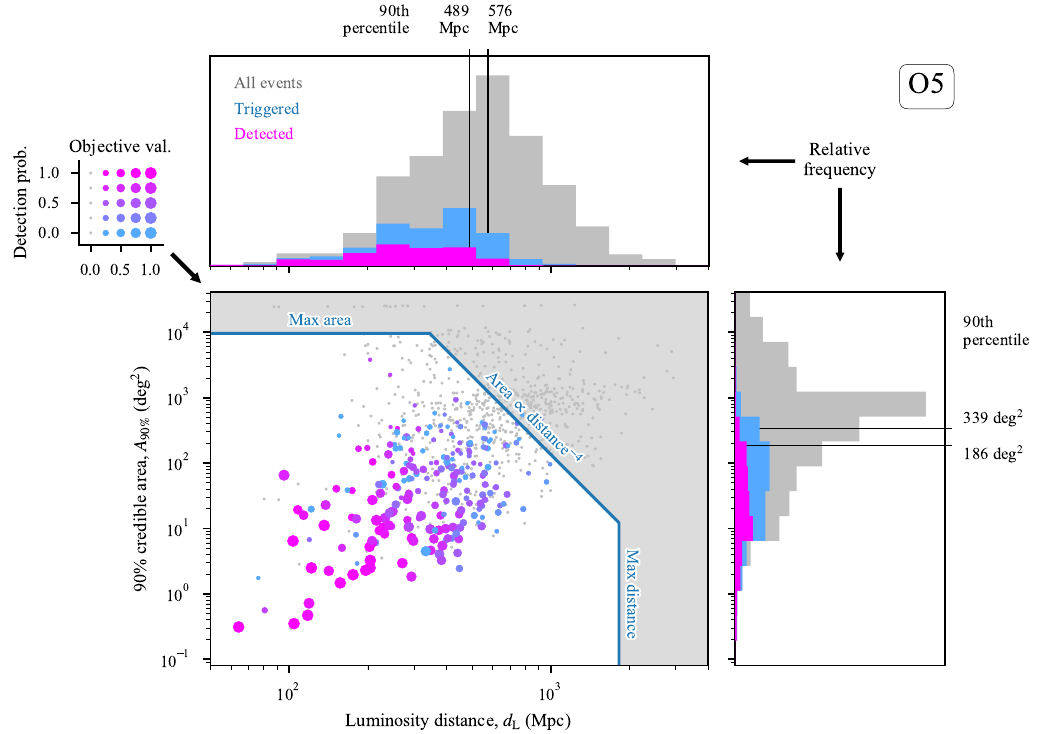}{0.8\textwidth}{}}
    \gridline{\fig{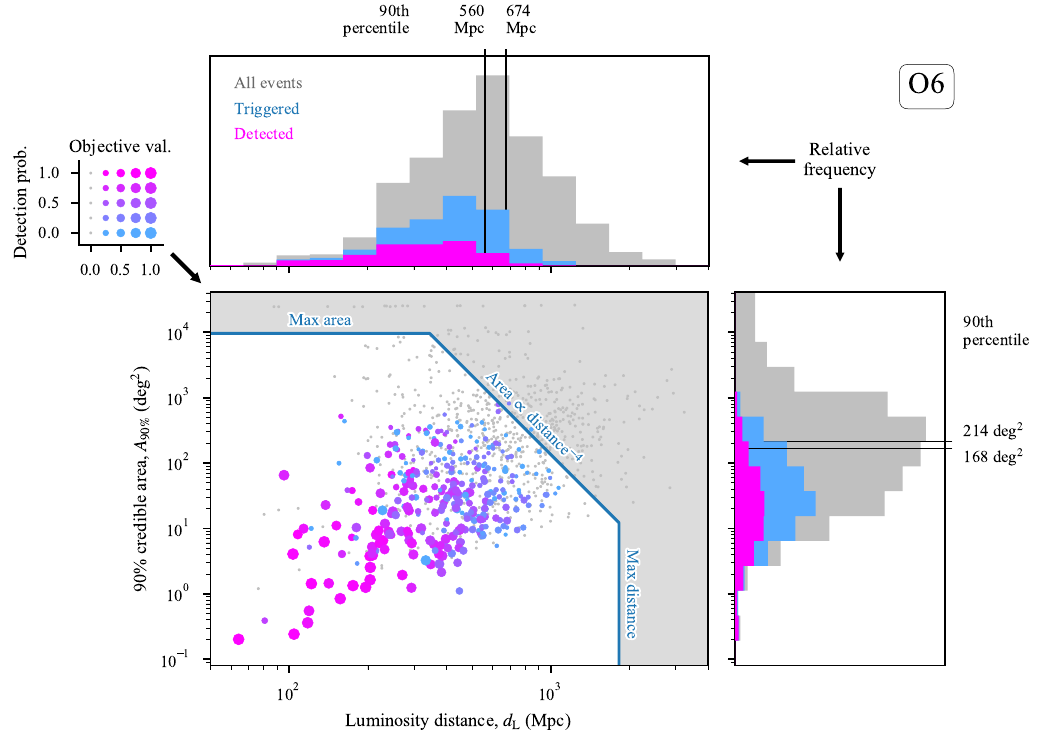}{0.8\textwidth}{}}
    \caption{\label{fig:area-distance}The distribution events selected for follow-up and detected for O5 and O6. Within each subfigure, there are three panels. \textbf{Bottom left panel:} scatter plot of 90\% credible area versus distance. Events that were selected for follow-up with \ac{UVEX} are represented by colored dots. The color of the dot represents the detection probability and the area of the dot represents the scheduler objective value. Events that were not selected for follow-up are marked with gray dots. The blue boundary represents the analytical predictor of the detection threshold given by Eqs.~(\ref{eq:threshold-distance},~\ref{eq:threshold-area},~\ref{eq:threshold-area-distance}). \textbf{Upper panel:} distribution of luminosity distance, with 90th percentiles marked. \textbf{Right panel:} distribution of 90\% credible area.}
\end{figure*}

In Fig.~\ref{fig:area-distance}, we show both the scheduler objective value and the detection probability on a scatter plot of the 90\% credible area and the luminosity distance of events. As we would expect, both the objective value and detection probability increase as the area and distance of the events decrease.

\begin{deluxetable*}{ccrrrrrrcc}
    \tablecaption{\label{tab:events}Simulated Events}
    \tablehead{
        & & \multicolumn2c{Source Frame Masses} & \multicolumn3c{True Position} & & \\
        \colhead{Run} & \colhead{Event ID} & \colhead{$m_1 / M_\odot$} & \colhead{$m_2 / M_\odot$} & \colhead{$\alpha$/deg} & \colhead{$\delta$/deg} & \colhead{$d_\mathrm{L}$/Mpc} & \colhead{$A_{90\%}$ / deg$^2$} & \colhead{Objective value\tablenotemark{\scriptsize{a}}} & \colhead{Detection prob.\tablenotemark{\scriptsize{b}}}
    }
    \startdata
    \input{tables/events.tex}
    \dots & \dots & \dots & \dots & \dots & \dots & \dots & \dots & \dots & \dots
    \enddata
    \tablenotetext{a}{The \ac{MILP} solver is configured to stop early if it can prove that the objective value is less than 0.1.}
    \tablenotetext{b}{Events for which the objective value is less than 0.1 are not scheduled for follow-up, so they have a detection probability of zero.}
    \tablecomments{Table~\ref{tab:events} is published in its entirety in a machine-readable format. A portion is shown here for guidance regarding its form and content.}
\end{deluxetable*}

In Table~\ref{tab:selected-detected}, we list the expected numbers of events selected for follow-up and the number of \acp{KN} detected over 1.5~yr of observation. The number of events selected is simply the number of observing plans with objective value $\geq 0.1$. The expected number of \acp{KN} detected is the sum of the detection probabilities of all of the events.

As in \citet{2022ApJ...924...54P} and in \citet{2023ApJ...958..158K}, this table gives central 90\% credible intervals about the median, incorporating a lognormal uncertainty in the \ac{BNS} merger rate of $210_{-120}^{240}$~Gpc$^{-3}$~yr$^{-1}$ and Poissonian variation in the number of events over a finite time duration.

If instead of optimizing for a detection in both filters, we optimize for a detection in \emph{at least one filter}, then both the number of events selected and the number of events detected approximately double.

\begin{deluxetable}{lcc}
    \tablecaption{\label{tab:selected-detected}Expected Number of Events}
    \tablehead{& O5 & O6}
    \startdata
    \input{tables/selected-detected.tex}
    \enddata
\end{deluxetable}

\section{Conclusion}

Unfortunately, direct comparisons of the rates in Table~\ref{tab:selected-detected} to \citet{2025PASP..137e4101C}, which used \texttt{dorado-scheduling} and a fixed exposure time for each event, are not meaningful because we made many refinements to the starting assumptions. The following change would lead to a higher detection rate:
\begin{itemize}
    \item We used a slightly brighter distribution of absolute magnitudes that is consistent with the stated \ac{KN} model ranges in \citet{2021arXiv211115608K}.
\end{itemize}
The following change would lead to a lower detection rate:
\begin{itemize}
    \item We used a much denser and more heavily overlapping grid of fields in order to ensure complete coverage of the sky at all roll angles.
\end{itemize}
Our trigger and detection rates in Table~\ref{tab:selected-detected} are comparable with the most optimistic estimates in \citet{2025PASP..137e4101C}. Although by itself this comparison is not informative on the merits of these two strategies, it is a positive result for the prospects of \ac{KN} detection with \ac{UVEX}.

\begin{figure}
    \includegraphics[width=\columnwidth]{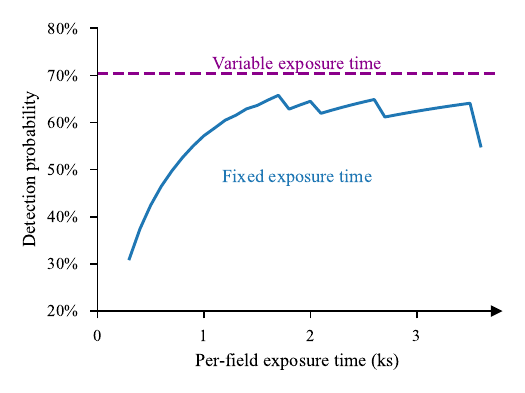}
    \caption{\label{fig:prob-exptime}Detection probability vs. exposure time for a single example event, resulting from the fixed exposure time strategy (see Section~\ref{sec:fixed-exptime}). The horizontal dashed line shows the detection probability resulting from the variable exposure time strategy incorporating a distribution of absolute magnitudes (see Section~\ref{sec:absmag-distn}). The variable exposure time strategy is more likely to detect the counterpart than the fixed exposure time strategy, for any exposure time.}
\end{figure}

More quantitatively, we can show that the dynamic exposure time strategy results in a higher detection probability than any given fixed exposure time for a given event. Fig.~\ref{fig:prob-exptime} shows a typical example. This is a plot of detection probability (without knowledge of the position of the source) versus exposure time for observing plans for a single event generated using the formulation from Section~\ref{sec:fixed-exptime}. The detection probability for the dynamic exposure time strategy from Section~\ref{sec:absmag-distn} is shown as a horizontal dashed line.

This novel dynamic exposure time observing strategy is promising for \ac{GW} follow-up with any large \ac{FOV} imaging instrument. Although we have focused on \ac{UVEX}, we encourage \ac{ULTRASAT} to try it out using the preliminary support for that mission that we have added. An \ac{M4OPT} command line option \texttt{--mission} which can take the values \texttt{uvex} or \texttt{ultrasat} selects all of the parameters described in Sections~\ref{sec:etc} and~\ref{sec:parameters} (\ac{ETC} model, bandpass, field of regard, \ac{FOV}, slew speed, field grid, and so on) between those that are appropriate for \ac{UVEX} or \ac{ULTRASAT} respectively.

The \ac{M4OPT} model itself is not specific to space-based observations. With additional \ac{FOR} and foreground models it could be extended to ground-based telescopes or even heterogeneous combinations of ground and space telescopes. In the future, we plan to extend it to \ac{ZTF}, Vera Rubin Observatory, and other ground-based facilities. Ultimately, our vision is for \ac{M4OPT} to evolve into a scheduling toolkit featuring (1) a composable library of observing constraints inspired by the interface of the \texttt{astroplan} package \citep{2018AJ....155..128M}, (2) uniform support for observers on the ground and in space, and (3) a globally optimizing scheduler.

When we constructed the luminosity function for the scheduler, we considered two \ac{KN} models that are broadly consistent with GW170817 and plausible for \ac{BNS} mergers: the jet-powered model of \citet{2018ApJ...855..103P} and the radioactively powered model of \citet{2020ApJ...891..152H}. However, the \ac{KN} emission is likely to be qualitatively different and more varied for \ac{NSBH} mergers. Indeed, the masses and spins of the original compact objects, which we are able to measure with \ac{GW} observations, can help us predict the nature of the remnant compact object and the \ac{KN} emission \citep{2019ApJ...880L..15M}. If \ac{GW} alerts included estimates of the masses, then we could use that information to taylor the times and depths of observations for the expected emission. Since the source lumninosity function in \ac{M4OPT} is configurable, in a future work we could evaluate mass-dependent observing strategies.

The scheduler currently considers a distribution of source absolute magnitudes, but not time variability of the source. This approach may be sufficient to optimize for obtaining a detection near peak, but is probably not sufficient to ensure a well-sampled light curve at later times. In a future work, we may add a light curve model which would allow the scheduler to automatically plan longer exposures before and after the predicted peak time of the \ac{KN} emission.

Our probabilistic approach to scheduling and triggering follow-up observations can in principle allow for a more detailed treatment of the selection effects incurred during \ac{GW} follow-up campaigns and astrophysical inference from the population of \acp{KN} observed by missions like \ac{UVEX}. Developing a framework for \ac{KN} inference with \ac{UVEX} observations triggered by this method is an exciting direction of future work.

Some practical implementation details remain as future work to incorporate \ac{M4OPT} into the operations of a real observatory. For example:
\begin{itemize}
\item Our slew time constraints must conservatively model the dynamics of the actual spacecraft, including differing angular rates about different axes and keep-out constraints that must hold through the entire slew trajectory.
\item We must be able to retrieve the spacecraft ephemeris from a live data source rather than evaluating it from a predefined orbital model.
\item We must incorporate pre-defined segments of observing dead time for housekeeping events such as pre-planned ground contacts and momentum dumping maneuvers.
\end{itemize}

Like the \ac{ZTF} scheduler \citep{2019PASP..131f8003B} and the Vera Rubin Observatory scheduler \citep{2019AJ....157..151N}, we see \ac{M4OPT} as both a core component of the ground software system that enables fully autonomous observation planning and execution, and also an approachable and well-documented part of the guest observer science tools for a mission. Its permissive open-source licensing and open development model, free from the encumbrances of \ac{NOSA}, are essential to those goals.

We hope to build from it a community tool that serves many missions from the concept phase all the way through operations. Also, integration of \ac{M4OPT} with community follow-up platforms such as the Gravitational Wave Treasure Map \citep{2020ApJ...894..127W} could open possibilities for more effective and more meaningfully coordinated follow-up observations by multiple independently operated facilities.

\begin{acknowledgments}
This work was performed in part at the Aspen Center for Physics, which is supported by \ac{NSF} grant PHY-2210452.

This work used Expanse at \ac{SDSC} and Delta at \ac{NCSA} through allocation AST200029, ``Towards a complete catalog of variable sources to support efficient searches for compact binary mergers and their products,'' from the \ac{ACCESS} program, which is supported by \ac{NSF} grants~\#2138259, \#2138286, \#2138307, \#2137603, and \#2138296.

S.C.L. and M.W.C. acknowledge support from \ac{NSF} with grant Nos. PHY-2308862 and PHY-2117997.

We thank Steve Crawford at NASA Headquarters for help and support with the NASA software release process.

The code, data, and software environment to reproduce the figures and tables in this paper are available from Zenodo \citep{singer_2025_15176276}.

This is LIGO document P2500008-v3.
\end{acknowledgments}

\vspace{5mm}
\software{
    \texttt{astropy} \citep{2013A&A...558A..33A,2018AJ....156..123A},
    \texttt{astroquery} \citep{2019AJ....157...98G},
    \texttt{dust\_extinction} \citep{2024JOSS....9.7023G},
    \texttt{dustmaps} \citep{2018JOSS....3..695M},
    \texttt{healpix} \citep{2005ApJ...622..759G},
    \texttt{healpy} \citep{2019JOSS....4.1298Z},
    \texttt{ligo.skymap} \citep{2016PhRvD..93b4013S,2016ApJ...829L..15S,2016ApJS..226...10S},
    \texttt{matplotlib} \citep{2007CSE.....9...90H},
    \texttt{m4opt} \citep{singer_2025_15169912},
    \texttt{numpy} \citep{harris2020array},
    \texttt{regions} \citep{larry_bradley_2022_6374572},
    \texttt{scipy} \citep{2020NatMe..17..261V},
    \texttt{spiceypy} \citep{2020JOSS....5.2050A},
    \texttt{sympy} \citep{10.7717/peerj-cs.103},
    \texttt{synphot} \citep{2018ascl.soft11001S}}

\appendix

\section{MILP Problem Details}
\label{sec:details}

In this appendix, we develop an \ac{MILP} formulation of the problem of planning and scheduling \ac{GW} \ac{ToO} observations on a large \ac{FOV} telescope. In Section~\ref{sec:fixed-exptime}, we start with a simple but realistic problem in which we require multiple visits of each selected field that must all obey the \ac{FOR} and slew speed limitations of the telescope. In Sections~\ref{sec:variable-exptime} and \ref{sec:absmag-distn}, we add progressively more detail, finally arriving at the full dynamic exposure time problem. All three formulations from Sections~\ref{sec:fixed-exptime}, \ref{sec:variable-exptime}, and \ref{sec:absmag-distn} are implemented in \ac{M4OPT}.

\subsection{Problem 1: Fixed Exposure Time}
\label{sec:fixed-exptime}

We receive a \ac{HEALPix} probability sky map that describes the probability distribution of the true but unknown position of a target of interest as a function of position on the sky. There is a delay between the time that the event occurred and when we can start observations due to the time it takes to uplink commands to the spacecraft, and there is a deadline by which we must complete our observations.

Our telescope can observe any of a set of $n_J$ fields at predetermined sky locations in order to tile the sky map. For each field that we select, our telescope must visit the field at least $n_K$ times. We have a cadence requirement that each visit of a given field must occur at least a time $\gamma$ after the previous visit. Multiple visits with a minimum cadence are essential in short timescale transient searches to rule out moving solar system objects, which otherwise are a major contaminant.

Every visit takes a certain amount of exposure time, and it takes a known amount of time to slew between different fields. We may only a visit a field when it is within the \ac{FOR}, the region that constrains where the telescope may point at any given instant of time.

\subsubsection{Data Preparation}

\begin{enumerate}
    \item Construct a discrete 1D grid of times that stretch from the delayed start of observations up to the deadline.
    \item Propagate the orbit of the spacecraft to calculate the position of the spacecraft at each time step.
    \item For each field and each time step, test whether the field is within the instantaneous \ac{FOR}, creating an observability bit map.
    \item Transform the observability bit map into a list of time segments during which each field is observable.
    \item Discard segments shorter than the exposure time.
    \item Discard fields that have no observable segments.
    \item For each field, find the \ac{HEALPix} pixel indices that are within the field's footprint.
    \item Select the 50 fields that contain the greatest probability, summed over the respective \ac{HEALPix} pixels.
    \item Discard pixels that are not contained in any field.
    \item Calculate the slew times between all pairs of distinct fields.
\end{enumerate}

\subsubsection{Problem Setup}

\begin{alignat*}{4}
\intertext{\it Index sets ---}
    &\text{pixels}&
        I =& \{0, 1, \dots, n_I - 1\} \\
    &\text{fields}&
        J =& \{0, 1, \dots, n_J - 1\} \\
    &\text{visits}&
        K =& \{0, 1, \dots, n_K - 1\} \\
    &\text{observable segments}&
        (M_j =& \{0, 1, \dots, {n_M}_j\})_{j \in J} \\
    &\text{fields containing pixel $i$}&
        (J_i =& \{0, 1, \dots, {n_J}_i\})_{i \in I} \\
\intertext{\it Parameters ---}
    &\text{probability of pixel $i$}&
        (\rho_i&)_{i \in I} \\
    &\text{slew time from field $j$ to $j^\prime$}&
        (\sigma_{jj^\prime}&)_{j \in J, j^\prime \in J} \\
    &\text{start times of segments}&
        (\alpha_{jm}&)_{j \in J, m \in M} \\
    &\text{end times of segments}&
        (\omega_{jm}&)_{j \in J, m \in M} \\
    &\text{exposure time}&
        \epsilon& \\
    &\text{cadence, time between visits}&
        \gamma& \\
    &\text{delay}&
        \beta& \\
    &\text{deadline}&
        \delta& \\
\intertext{\it Binary decision variables ---}
    &\text{pixel $i$ is in any selected field}&
        (p_i&)_{i \in I} \\
    &\text{field $j$ is selected}&
        (r_j&)_{j \in J} \\
    &\text{field $j$ visit $k$ is in segment $m$}&
        (s_{jkm}&)_{j \in J, k \in K, m \in M \mid {n_M}_j > 1} \\
\intertext{\it Continuous decision variables ---}
    &\text{mid time of field $j$ visit $k$}& (t_{jk}&)_{j \in J, k \in K}
\end{alignat*}

\subsubsection{Constraints}

\paragraph{Containment}
Only count pixels that are in one or more selected fields.
\begin{equation}
    \label{eq:fixed-exptime-constraint-containment}
    \forall i :\quad p_i \leq \sum_{j \in J_i} r_j
\end{equation}

\paragraph{Cadence}
If a field is selected for observation, then enforce a minimum time between visits.
\begin{equation}
    \label{eq:fixed-exptime-constraint-cadence}
    \forall k > 1 ,\; j :\quad t_{jk} - t_{j,k-1} \geq (\epsilon + \gamma) r_j
\end{equation}

\paragraph{No overlap}
Observations cannot overlap in time; they must be separated by at least the exposure time plus the slew time.
\begin{multline}
    \label{eq:fixed-exptime-constraint-no-overlap}
    \forall j^\prime > j,\; k ,\; k^\prime : \\ \left|t_{jk} - t_{j^\prime k^\prime}\right|  \geq \left(\sigma_{jj^\prime} + \epsilon\right) \left( r_j + r_{j^\prime} - 1\right)
\end{multline}

\paragraph{\ac{FOR}}
An observation of a field can only occur while the coordinates of the field are within the \ac{FOR}. For fields that have one observable segment (${n_M}_j = 1$), this constraint is simply an inequality:
\begin{equation}
    \label{eq:fixed-exptime-constraint-for-one}
    \forall j ,\; k \;, m \mid {n_M}_j = 1 :\quad \alpha_{jm} + \epsilon / 2 \leq t_{jk} \leq \omega_{jm} - \epsilon / 2
\end{equation}
For fields that have more than one observable segment (${n_M}_j > 1$), we use the decision variable $s_{jkm}$ to determine which inequality is satisfied:
\begin{alignat}{2}
    \label{eq:fixed-exptime-constraint-for-many}
    \forall j ,\; k \;, &m \mid {n_M}_j > 1 : \nonumber \\
    &s_{jkm} = 1 \;\Rightarrow\; \alpha_{jm} + \epsilon / 2 \leq t_{jk} \leq \omega_{jm} - \epsilon / 2 \\
    &\sum_m s_{jkm} \geq 1
\end{alignat}

\subsubsection{Cuts}

\paragraph{Total exposure time}
Although it is implied by other constraints, the constraint that the total exposure time cannot exceed the total available time is found to speed up the search. We add it as a cut: an inequality that the \ac{MILP} may use to help guide its search but that is not checked when evaluating integer feasibility.
\begin{equation}
    \label{eq:fixed-exptime-cut-total-time}
    \sum_{j \in J} r_j \leq \frac{\delta - \beta}{\epsilon n_K}
\end{equation}

\subsubsection{Objective}
\label{sec:fixed-exptime-objective}

Maximize the sum of the probability of all of the pixels that are contained within selected fields:
\begin{equation}
    \label{eq:fixed-exptime-objective}
    \sum_{i \in I} \rho_i p_i
\end{equation}

\subsection{Problem 2: Variable Exposure Time}
\label{sec:variable-exptime}

In this variation, we have a sky map of the exposure time required to detect the source as a function of its position on the sky. We permit the exposure time to vary for each field. A given pixel counts toward the objective value only if the exposure time of a field that contains that pixel exceeds the pixel's exposure time.

\subsubsection{Problem Setup}
\label{sec:variable-exptime-problem-setup}

\begin{alignat*}{3}
\intertext{\it Additional parameters ---}
    &\text{min exposure time to detect a source in pixel $i$}\quad&
        (\epsilon_i)_{i \in I} \\
    &\text{min allowed exposure time}\quad&
        \epsilon_\mathrm{min} \\
    &\text{max allowed exposure time}\quad&
        \epsilon_\mathrm{max}
\intertext{\it Additional, semicontinuous decision variables ---}
    &\text{exposure time of field $j$}\quad& \\
    &\qquad\left(e_j\right)_{j \in J}, \forall j \in J : e_j = 0 \text{ or } \epsilon_\mathrm{min} \leq e_j \leq \epsilon_\mathrm{max}
\end{alignat*}

\subsubsection{Constraints}

The constraints are slightly different.

\paragraph{Depth}
Only count pixels that are observed to sufficient exposure time.
\begin{equation}
    \label{eq:variable-exptime-constraint-depth}
    \forall i \in I :\quad p_\mathrm{i} = 1 \Rightarrow \max_{j \in J_i} e_{j} \geq \epsilon_i
\end{equation}

\paragraph{Exposure time}
If a field's exposure time is nonzero, then it is selected for observation.
\begin{equation}
    \label{eq:variable-exptime-constraint-exptime}
    \forall j \in J :\quad \epsilon_\mathrm{max} r_j \geq e_\mathrm{j}
\end{equation}

\paragraph{Cadence}
This is similar to Eq.~(\ref{eq:fixed-exptime-constraint-cadence}), except that we replace the right-hand side of the inequality.
\begin{equation}
    \label{eq:variable-exptime-constraint-cadence}
    \forall k > 1 ,\; j :\quad t_{jk} - t_{j,k-1} \geq \gamma r_j + e_j
\end{equation}

\paragraph{No overlap}
This is also similar to Eq.~(\ref{eq:fixed-exptime-constraint-no-overlap}), except with a slightly different right-hand side.
\begin{multline}
    \label{eq:variable-exptime-constraint-no-overlap}
    \forall j^\prime > j ,\; k ,\; k^\prime : \\ \left|t_{jk} - t_{j^\prime k^\prime}\right|  \geq \sigma_{jj^\prime} \left( r_j + r_{j^\prime} - 1\right) + (e_j + e_\mathrm{j^\prime}) / 2
\end{multline}

\paragraph{\ac{FOR}}
This is similar to Eqs.~(\ref{eq:fixed-exptime-constraint-for-one},~\ref{eq:fixed-exptime-constraint-for-many}), except that we replace $\epsilon$ with $e_j$. For fields that have one observable segment:
\begin{multline}
    \label{eq:variable-exptime-constraint-for-one}
    \forall j ,\; k \;, m \mid {n_M}_j = 1 : \\ \alpha_{jm} + e_j / 2 \leq t_{jk} \leq \omega_{jm} - e_j / 2
\end{multline}
For fields that have more than one observable segment:
\begin{alignat}{2}
    \label{eq:variable-exptime-constraint-for-many}
    \forall j ,\; &k \;, m \mid {n_M}_j > 1 : \nonumber \\
    &s_{jkm} = 1 \;\Rightarrow\; \alpha_{jm} + e_j / 2 \leq t_{jk} \leq \omega_{jm} - e_j / 2 \\
    &\sum_m s_{jkm} \geq 1
\end{alignat}

\subsubsection{Cuts}

\paragraph{Total exposure time}
Replace Eq.~(\ref{eq:fixed-exptime-cut-total-time}) with:
\begin{eqnarray}
    \label{eq:variable-exptime-cut-total-time}
    \sum_{j \in J} r_j &\leq& \frac{\delta - \beta}{\epsilon_\mathrm{min} n_K} \\
    \sum_{j \in J} e_j &\leq& \frac{\delta - \beta}{n_K}
\end{eqnarray}

\subsubsection{Objective}

Same as in Section~\ref{sec:fixed-exptime-objective} above.

\subsection{Problem 3: Variable Exposure Time with Prior Distribution of Absolute Magnitude}
\label{sec:absmag-distn}

In this variation, we do not know the precise absolute magnitude $X$ of the source. In the case of \acp{KN}, our prior knowledge about the absolute magnitude is scant; for the sake of mathematical convenience, we assume that the absolute magnitude has a normal distribution, $X \sim \mathcal{N}[\mu_X, \sigma_X]$. We need to compute the distribution of \textit{apparent} magnitudes $x$ in order to determine the probability of detection as a function of exposure time for each pixel.

Gravitational-wave sky maps provide the posterior distribution of distance, as a parametric ansatz distribution \citep{2016ApJ...829L..15S,2016ApJS..226...10S},
$$
    p(r) = \frac{N}{\sqrt{2 \pi}\sigma} \exp\left[-\frac{1}{2}\left(\frac{r - \mu}{\sigma}\right)^2\right] r^2,
$$
with the location parameter $\mu$, scale parameter $\sigma$, and normalization $N$ tabulated for each pixel. This is an inconvenient distribution for integration, so instead we construct a log-normal distance distribution with the same mean and standard deviation as the ansatz distribution.

We calculate the mean $m$ and standard deviation $s$ from $\mu$ and $\sigma$ using the function \texttt{parameters\_to\_moments} from \texttt{ligo.skymap}.\footnote{\url{https://lscsoft.docs.ligo.org/ligo.skymap/distance/}} Then, the location and scale parameters of the log-normal distribution are given by
\begin{eqnarray}
    \label{eq:log-distance-parameters}
    \mu_{\ln r} &=& \ln m - \frac{1}{2} \ln \left(1 + \frac{s^2}{m^2}\right) \\
    {\sigma_{\ln r}}^2 &=& \ln \left(1 + \frac{s^2}{m^2}\right).
\end{eqnarray}

The logarithm of the distance then has the distribution $\ln r \sim \mathcal{N}[\mu_{\ln r}, \sigma_{\ln r}]$. The apparent magnitude is related to the absolute magnitude through $x = X + 5 \log_{10} r + 25$, assuming that $r$ is in the units of Mpc. Therefore the apparent magnitude has the distribution $x \sim \mathcal{N}[\mu_x, \sigma_x]$, with
\begin{eqnarray}
    \label{eq:appmag-parameters}
    \mu_x &=& \mu_X + \left(\frac{5}{\ln 10}\right) \mu_{\ln r} + 25 \\
    {\sigma_x}^2 &=& {\sigma_{X}}^2 + \left(\frac{5}{\ln 10}\right)^2 {\sigma_{\ln r}}^2.
\end{eqnarray}

\begin{figure}
    \includegraphics[width=\columnwidth]{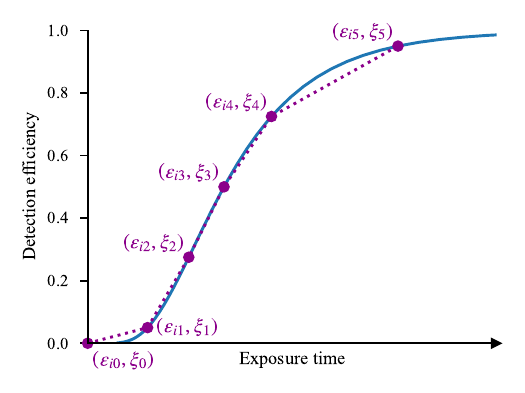}
    \caption{\label{fig:piecewise-linear-exptime}Piecewise linear approximation of the detection efficiency for a given pixel. In this example, we have assumed that the exposure time is inversely proportional to the square root of the flux, valid for sky background dominated imaging.}
\end{figure}

With this Gaussian distribution of apparent magnitudes, we can now calculate the detection efficiency for each pixel: the probability that we detect the source assuming that the source is in that pixel, as a function of exposure time. For the purpose of implementation of this function in a \ac{MILP}, we approximate it with a piecewise linear function, as illustrated in Fig~\ref{fig:piecewise-linear-exptime}.

\subsubsection{Additional Data Preparation}

\begin{enumerate}
    \item Use the function \texttt{parameters\_to\_moments} and Eqs.~(\ref{eq:log-distance-parameters},~\ref{eq:appmag-parameters}) to calculate the mean and standard deviation of the apparent magnitude in each pixel.
    \item Select the desired quantiles for the piecewise linear approximation of the detection efficiency curve: for example, $(0, 0.05 , 0.275, 0.5  , 0.725, 0.95)$. For each pixel, calculate the exposure time required to achieve the specified detection efficiency.
\end{enumerate}

\subsubsection{Problem Setup}

\begin{alignat*}{3}
\intertext{\it Additional index sets ---}
    &\text{indices of quantiles}&\quad
        N = \{0, 1, \dots, n_N\} \\
\intertext{\it Additional parameters ---}
    &\text{quantiles of detection efficiency}&\quad
        (\xi_n)_{n \in N} \\
    &\text{exposure time of quantiles}&\quad
        (\epsilon_{in})_{i \in I, n \in N} \\
    &\text{piecewise linear functions}&\quad
        (f_i: \mathbb{R}_{\geq 0} \rightarrow [0, 1])_{i \in I} \\
\intertext{\it Additional, continuous decision variables ---}
    &\text{change from binary to continuous}&\quad (p_i)_{i \in I}
\end{alignat*}

\subsubsection{Additional Constraints}

\paragraph{Depth}
Replace Eq.~\ref{eq:variable-exptime-constraint-depth} with:
$$
    \forall i \in I :\quad \max_{j \in J_i} e_{j} \geq f_i(p_\mathrm{i})
$$

\subsubsection{Objective}

Same as in Section~\ref{sec:fixed-exptime-objective} above.

\bibliography{m4opt}{}
\bibliographystyle{aasjournal}

\end{document}

%% file: tables/example.tex
10:23:42 & 325 & observe & -183 & 142 & -50 & 31.6571 & -23.5560 & -138 & 0.04 & 255 & 24.07 & 0.07 & 27.36 & 0.01 \\
10:29:06 & 120 & slew & --- & --- & --- & --- & --- & --- & --- & --- & --- & --- & --- & --- \\
10:31:06 & 767 & observe & -179 & 148 & -50 & 32.4155 & -28.4429 & -141 & 0.15 & 278 & 24.83 & 0.07 & 27.48 & 0.08 \\
10:43:53 & 127 & slew & --- & --- & --- & --- & --- & --- & --- & --- & --- & --- & --- & --- \\
10:46:00 & 1348 & observe & -168 & 160 & -50 & 36.9122 & -33.0462 & -146 & 0.21 & 320 & 25.29 & 0.11 & 27.54 & 0.11 \\
11:08:28 & 107 & slew & --- & --- & --- & --- & --- & --- & --- & --- & --- & --- & --- & --- \\
11:10:15 & 1529 & observe & -149 & 179 & -50 & 39.3087 & -35.3375 & -149 & 0.11 & 345 & 25.39 & 0.16 & 27.57 & 0.05 \\
11:35:44 & 127 & slew & --- & --- & --- & --- & --- & --- & --- & --- & --- & --- & --- & --- \\
11:37:51 & 3006 & observe & -125 & 197 & -50 & 34.6192 & -30.7459 & -143 & 0.27 & 296 & 25.84 & 0.10 & 27.51 & 0.19 \\
12:27:57 & 120 & slew & --- & --- & --- & --- & --- & --- & --- & --- & --- & --- & --- & --- \\
12:29:57 & 2166 & observe & -75 & 223 & -50 & 33.7574 & -25.7998 & -141 & 0.16 & 269 & 25.58 & 0.07 & 27.40 & 0.11 \\
13:06:04 & 108 & slew & --- & --- & --- & --- & --- & --- & --- & --- & --- & --- & --- & --- \\
13:07:51 & 924 & observe & -34 & 234 & -50 & 35.9256 & -28.0474 & -144 & 0.17 & 284 & 24.97 & 0.08 & 27.43 & 0.09 \\
13:23:15 & 107 & slew & --- & --- & --- & --- & --- & --- & --- & --- & --- & --- & --- & --- \\
13:25:02 & 767 & observe & -15 & 236 & -50 & 32.4155 & -28.4429 & -141 & 0.15 & 278 & 24.83 & 0.07 & 27.48 & 0.08 \\
13:37:49 & 120 & slew & --- & --- & --- & --- & --- & --- & --- & --- & --- & --- & --- & --- \\
13:39:49 & 325 & observe & 1 & 237 & -50 & 31.6571 & -23.5560 & -138 & 0.04 & 255 & 24.07 & 0.07 & 27.36 & 0.01 \\
13:45:14 & 108 & slew & --- & --- & --- & --- & --- & --- & --- & --- & --- & --- & --- & --- \\
13:47:01 & 2166 & observe & 9 & 237 & -50 & 33.7574 & -25.7998 & -141 & 0.16 & 269 & 25.58 & 0.07 & 27.40 & 0.11 \\
14:23:08 & 108 & slew & --- & --- & --- & --- & --- & --- & --- & --- & --- & --- & --- & --- \\
14:24:55 & 924 & observe & 51 & 233 & -50 & 35.9256 & -28.0474 & -144 & 0.17 & 284 & 24.97 & 0.08 & 27.43 & 0.09 \\
14:40:19 & 104 & slew & --- & --- & --- & --- & --- & --- & --- & --- & --- & --- & --- & --- \\
14:42:04 & 3006 & observe & 70 & 229 & -51 & 34.6192 & -30.7459 & -143 & 0.27 & 296 & 25.84 & 0.10 & 27.51 & 0.19 \\
15:32:10 & 107 & slew & --- & --- & --- & --- & --- & --- & --- & --- & --- & --- & --- & --- \\
15:33:57 & 1348 & observe & 123 & 207 & -51 & 36.9122 & -33.0462 & -146 & 0.21 & 320 & 25.29 & 0.11 & 27.55 & 0.11 \\
15:56:25 & 107 & slew & --- & --- & --- & --- & --- & --- & --- & --- & --- & --- & --- & --- \\
15:58:12 & 1529 & observe & 146 & 192 & -51 & 39.3087 & -35.3375 & -149 & 0.11 & 345 & 25.39 & 0.16 & 27.57 & 0.05

%% file: tables/events.tex
O5 & 14 & 1.976 & 1.597 & 217.5627 & +50.1543 & 204 & 3 & \phantom{$<$}0.95 & 0.92 \\
O5 & 27 & 1.605 & 1.477 & 9.8234 & +23.2808 & 685 & 571 & $<$0.10 & --- \\
O5 & 30 & 1.318 & 1.280 & 337.1588 & -36.0109 & 253 & 192 & \phantom{$<$}0.16 & 0.00 \\
O5 & 41 & 1.979 & 1.772 & 194.5206 & -19.3125 & 488 & 129 & \phantom{$<$}0.16 & 0.32 \\
O5 & 46 & 6.405 & 2.379 & 139.4866 & -25.4656 & 1277 & 766 & $<$0.10 & --- \\
O5 & 52 & 6.754 & 2.218 & 335.1421 & -3.1377 & 1457 & 90 & $<$0.10 & --- \\
O5 & 54 & 1.961 & 1.583 & 131.7071 & -38.2637 & 679 & 1143 & $<$0.10 & --- \\
O5 & 60 & 25.417 & 2.561 & 91.4538 & +78.0966 & 800 & 4579 & $<$0.10 & --- \\

%% file: tables/selected-detected.tex
Number of events selected & $29_{-18}^{+39}$ & $43_{-26}^{+56}$ \\
Number of events detected & $12_{-9}^{+18}$ & $17_{-11}^{+24}$